# Logical error rates for the surface code under a mixed coherent and stochastic circuit-level noise model inspired by trapped ions

Tyler LeBlond [1,*,†] Peter Groszkowski [2] Justin G. Lietz [2,‡] Christopher M. Seck [1] and Ryan S. Bennink [1]

[1] *Computational Sciences and Engineering Division, Oak Ridge National Laboratory, Oak Ridge, Tennessee 37831, USA*

[2] *National Center for Computational Sciences, Oak Ridge National Laboratory, Oak Ridge, Tennessee 37831, USA*

With fault-tolerant quantum computing (FTQC) on the horizon, it is critical to understand sources of logical errors in plausible hardware implementations of quantum error-correcting codes. Detailed error modeling of computational instructions on particular FTQC architectures will enable the better prediction of error propagation in FT-encoded quantum circuits while revealing where greater attention is needed in hardware design. In this work, we consider logical error rates for the surface code implemented on a hypothetical grid-based trapped-ion quantum charge-coupled device architecture. Specifically, we construct logical channels for the idling surface code and examine its diamond error under a mixed coherent and stochastic circuit-level noise model inspired by trapped ions. We include the coherent dephasing noise that is known to accumulate during physical qubit idling and transport in these systems, determining idling and transport durations using the time-resolved output of an open-source trapped-ion surface code compiler. To estimate expectation values of logical Pauli observables following hardware circuits containing non-Clifford sources of noise, we utilize a Monte Carlo technique to sample from an underlying quasiprobability distribution of Clifford circuits that we independently simulate in a phase-sensitive fashion. We verify error suppression up to code distance $d = 11$ at coherent dephasing rates near and below those of current-generation trapped-ion quantum computers and find that logical error rates align with those of analogous fully stochastic simulations in this regime. Exploring higher dephasing rates at $d = 3$–$5$, we find evidence for growing coherent rotations about all three logical Pauli axes, increased diagonal logical error process matrix elements relative to those of stochastic simulations, and a reduced dephasing rate threshold. Overall, our work paves a way toward realistic hardware emulation of small fault-tolerant quantum processes, e.g., members of an FTQC instruction set.

## I. INTRODUCTION

Recently, experimental demonstrations of important early milestones in quantum error correction (QEC) such as beyond-break-even logical error rates, subthreshold logical error rate scaling, fault-tolerant entangling gates and logical qubit teleportation, logical magic state distillation, and beyond-break-even non-Clifford gate infidelities have ushered in the early fault-tolerant era of quantum computing [1–9]. Even so, consensus has not been achieved on which combinations of quantum computing modality and quantum error-correcting code (QECC) are most promising in the industry. The surface code garnered a substantial amount of early attention in fault-tolerant quantum computing (FTQC) architecture studies due to its high threshold under circuit-level noise and nearest-neighbor connectivity [10–12], but its high-qubit overhead and vanishing encoding rate have led many to search for protocols involving constant-rate quantum low-density parity check (qLDPC) or concatenated codes [13–16]. The surface code remains a leading QECC for planar architectures, e.g., those based upon superconducting circuits and Majorana zero modes [1,2,17,18], while architectures with all-to-all connectivity such as those based on neutral atoms and trapped ions supply the freedom to explore more exotic schemes [5,9,14]. That said, the surface code sustains interest for such systems because it serves as a conservative baseline for performance and resource comparison [4,19–22] while schemes with higher encoding rates, which are still at a lower level of theoretical maturity, are being explored. For these reasons, it is timely to study the logical error performances of competitive FTQC architectures, and the surface code mapped to trapped-ion quantum charge-coupled device (QCCD) hardware is one important baseline to consider.

In this study, we focus our attention on one particular grid-based trapped-ion QCCD architecture proposal and surface code mapping for which open-source compilation software is available [21], which we call the trapped-ion surface code (TISC). Advanced compiler frameworks that have been proposed for this architecture [19,20] as well as recent advances in experimental trapped-ion quantum computing [23] make the TISC a realistic target for future FTQC demonstrations.

*Contact author: tyler.leblond@quantinuum.com

†Present address: Quantinuum, Broomfield, USA.

‡Present address: NVIDIA Corporation, Santa Clara, USA.

Our goal in the present work is to construct logical error channels and calculate logical error rates for the idling TISC under a noise model inspired by trapped ions in order to better understand the influence of physical coherent noise in this architecture.

Recent demonstrations of QEC protocols on current-generation trapped-ion QCCD systems have noted the influence of coherent dephasing noise on the results and/or have taken action to mitigate its effects [7,24–26]. Past theoretical work has shown that physical-layer coherent noise manifests in the logical (residual) error channel of an "idling" error-corrected logical qubit as coherent logical noise [27–29], which is expected to cause faster nonlinear growth of logical infidelity than would result from purely stochastic physical noise. Since conventional resource analysis frameworks rely on the approximation that total logical error rates accumulated during quantum circuit execution depend linearly on logical error rates per QEC cycle derived from stochastic Pauli noise simulations [30,31], these analyses may underestimate resource costs for quantum computing systems with physical-layer coherent noise, opening up questions about their scalability. Though it is expected that coherent logical noise will be suppressed with increasing code distance [27–29,32], it is unclear at which code distances it will be negligible in practical setups, or even whether this conclusion will hold under circuit-level coherent noise.

Aside from the translation from physical-layer coherent error to logical-layer coherent error (and the possible suppression of the latter with increasing code distance), it has been observed numerically that significant differences exist between logical error rates derived from simulations that treat coherent physical noise using its Pauli twirl vs those that do not [32–34]. For instance, the results in Ref. [32] appear to suggest that the ratio of the two quantities decreases exponentially with increasing code distance in the subthreshold scaling regime. These findings were corroborated by Ref. [35] in the presence of phenomenological readout errors, but the same has not been studied in the context of circuit-level coherent noise. For these reasons, concerns about coherent dephasing noise affecting the scalability of trapped-ion QCCD systems may have validity.

In this article, we explore these questions for a concrete set of architectural assumptions. In particular, we provide simulations of the TISC up to code distance $d = 11$ under a mixed coherent and stochastic circuit-level noise model that includes the accumulation of coherent dephasing error during physical qubit idling and movement. Using the Trapped-Ion Surface Code Compiler (TISCC) to obtain the durations of time elapsed between gates applied to physical qubits [21], we apply coherent $Z$-axis rotations before quantum gates and measurements with rotation angles that depend on these idling durations and the coherent dephasing rate. We sweep over the coherent dephasing rate near current experimental values while holding the strengths of other noise sources (which are all modeled as stochastic Pauli channels) constant.

Our main findings include that, under this noise model, the TISC is able to suppress logical error rates near current experimental coherent dephasing rates, with its logical error rates almost exactly following those obtained by simulating the same error model with all coherent errors Pauli twirled. This result indicates that there is a regime in which there is no benefit to utilizing methods such as dynamical decoupling to effectively Pauli twirl physical coherent noise. At higher dephasing rates, for which we obtained results at $d = 3-5$, we find growing logical coherent rotations about all three Pauli axes, as well as larger diagonal logical error process matrix elements compared with analogous fully stochastic simulations, leading to higher logical error rates and a reduced threshold dephasing rate.

Coherent noise simulations of QEC are typically considered to be out of reach numerically for practical code distances due to the noise-incorporated circuits being nonstabilizer, though some progress has recently been made in developing methods to this end [36]. Also, though a method for simulating coherent errors using a Majorana fermion mapping of the surface code has been developed [32], this technique has limitations; while it has recently been extended to accommodate measurement errors [35], it is unclear how it could be extended to accommodate circuit-level coherent errors of the kind we study in this work. Herein, we build upon the strategy of Ref. [37] to estimate logical Pauli expectation values of the TISC using Clifford decompositions of noisy hardware circuits and Monte Carlo sampling over the resulting quasiprobability distribution of Clifford circuits. Our simulation methodology, which we describe in detail in this paper, is based upon Bravyi's CH-form for stabilizer states and phase-sensitive stabilizer simulation method [38], but adapted to the setting where expectation values of logical Pauli observables are the desired output rather than the distribution of bit strings. Estimating logical Pauli observables for the idling TISC given an informationally complete set of logical input states enables us to estimate logical Pauli transfer matrix (PTM) elements, which in turn enables us to calculate logical error rates based on the diamond distance.

Although it is known that coherent error accumulates during ion cooling in trapped-ion QCCD architectures [39,40], we do not incorporate coherent error into these operations, as we do not treat ion cooling explicitly. Therefore, our simulations likely under-represent the physical coherent noise present in the TISC, and our results should be taken as indicative of qualitative trends rather than quantitative benchmarks. That said, it is expected that future-generation systems will pipeline operations within gate zones in such a way that time overhead due to cooling is effectively eliminated [41]. Future work should incorporate these architectural assumptions.

The remainder of this paper will be organized as follows. In Sec. II, we summarize the major contributions of this work. In Sec. III, we discuss our TISC mapping and the physical-layer error model we consider. In Sec. IV, we discuss our near-Clifford simulation methodology and computational workflow for extracting process matrices and logical error rates. In Sec. V, we discuss the results of our work. Finally, in Sec. VI we provide concluding remarks.

## II. MAJOR CONTRIBUTIONS

(i) We have emulated a realistic trapped-ion QCCD surface code mapping under a mixed coherent and stochastic circuit-level noise model. We include coherent dephasing errors during physical qubit idling and movement, which are believed to be a dominant noise source in trapped-ion QCCD architectures [24,25].

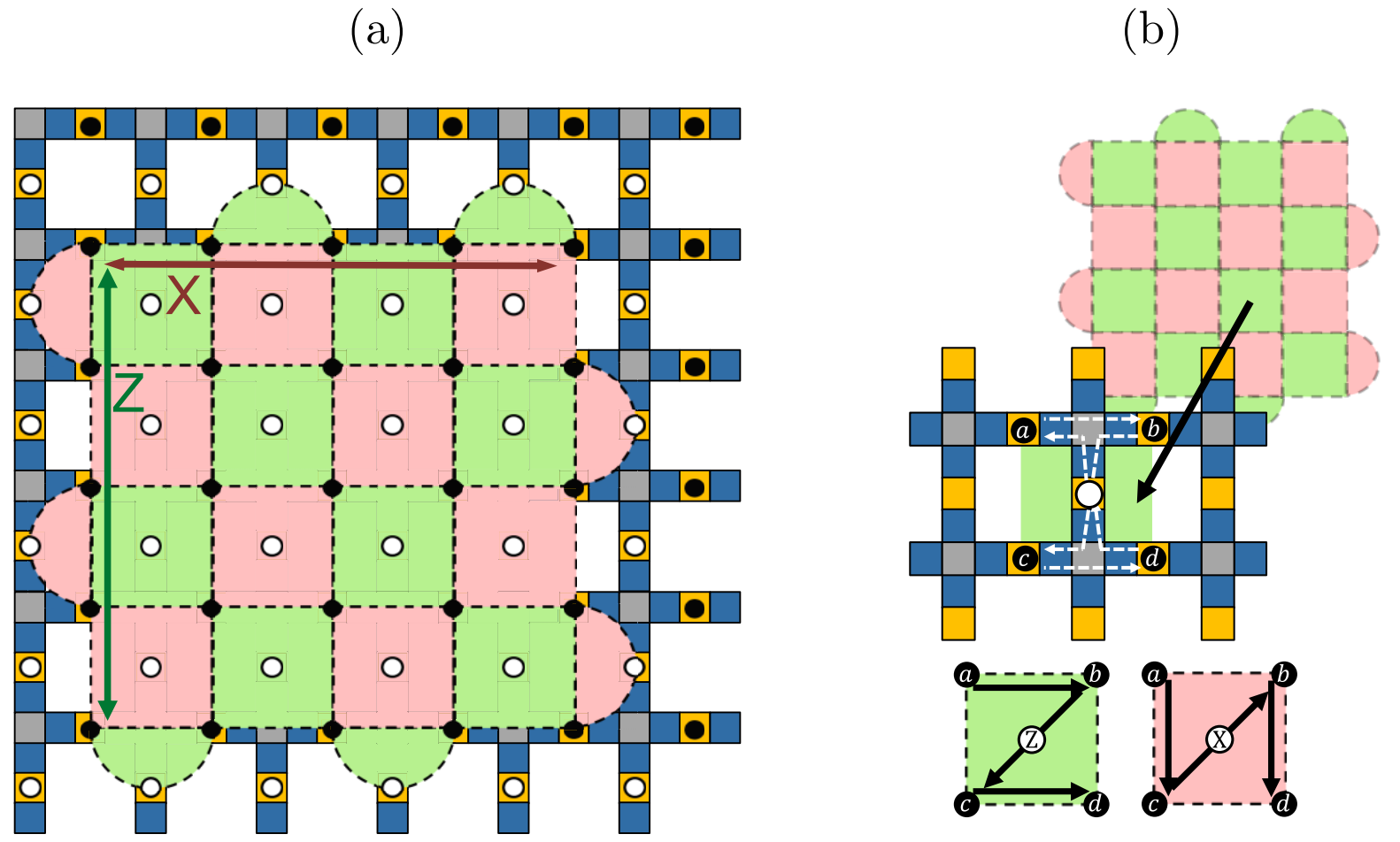

(c)

| Operation | Time (μs) |
|---|---|
| Initialization | 10 |
| Measurement | 120 |
| $R_X(\pi)$ | 10 |
| $R_X(\pi/2)$ | 10 |
| $R_Y(\pi)$ | 10 |
| $R_X(\pi/2)$ | 10 |
| $R_Z(\pi)$ | 3 |
| $R_Z(\pi/2)$ | 3 |
| $R_Z(\pi/4)$ | 3 |
| $R_{ZZ}(\pi/2)$ | 2000 |
| Move (Zone) | 5.25 |
| Move (Junction) | 105 |

FIG. 1. Adapted from Ref. [21]. (a) Pictorial representation of a surface code patch in what we call the "standard arrangement" of stabilizers, with $X$ stabilizers colored red (dark) and $Z$ stabilizers colored green (light). The patch is superimposed over a section of the trapped-ion hardware that we abstractly call a "logical tile," with memory sites colored dark blue, operations sites colored gold, and junction sites colored gray. Sites are shown occupied by data qubits (black circles) and measure qubits (white circles). (b) To interact, the measure qubit (white) moves adjacent to each data qubit (black) in a sequence specified by the measurement pattern, which is different for $X$ and $Z$ stabilizers. (c) The native trapped-ion instruction set implemented within TISCC, converted to the notation of this paper.

(ii) To the best of our knowledge, we provide the first study of logical error in the surface code under mixed coherent and stochastic circuit-level noise above $d = 5$. We provide results up to $d = 11$ within a regime of practical interest.

(iii) We numerically estimate logical process matrices for the surface code idling under such a noise model and find that, near the coherent dephasing rates of experimental systems, logical error rates are consistent with those of the analogous fully stochastic noise simulations up to $d = 11$.

(iv) At higher coherent dephasing rates, for which we obtained results at $d = 3$–$5$, we find evidence that physical coherent error manifests as logical coherent error, increases logical error rates, and reduces the dephasing rate threshold relative to analogous fully stochastic noise simulations.

(v) We reveal the methodology behind our Oak Ridge Quasi-Clifford Simulator (ORQCS), which enables us to estimate Pauli-basis expectation values for near-Clifford quantum circuits. We also provide a scaling analysis for the method and demonstrate empirically where its performance breaks down.

## III. TRAPPED-ION HARDWARE EMULATION

### A. Mapping the surface code to a trapped-ion QCCD hardware model

Several specialized mappings have been proposed for the surface code onto trapped-ion quantum computing architectures [21,42–45]. For those architectures consisting of linear chains, Trout *et al.* [43] presented an optimized mapping and detailed simulation of the distance three surface code, and very recently, Ye and Delfosse presented a mapping that optimized the number of ancilla qubits needed for stabilizer measurement up to distance seven [45]. Due to limitations on the number of ions that can be supported by a single chain, scalability to higher code distances requires the use of several ion chains together. In this vein, Li and Benjamin proposed a surface code mapping to a segmented linear chain and performed simulations demonstrating a threshold of 0.7% [44]. Unfortunately, the scalability of this scheme is also hindered by a syndrome extraction cycle time that is linear in the code distance [44]. Since the trapped-ion QCCD architecture has the potential for massive parallelism of operations, it presents a scalability advantage to the segmented linear chain [39,46,47].

We have previously proposed a mapping for a universal set of surface code logic operations onto a hypothetical grid-like trapped-ion QCCD architecture [21,48]. The mapping in Ref. [21] was given in order to underlay the TISCC, an open-source compiler capable of generating hardware circuits (including explicit ion shuttling instructions) for those operations. The output circuits of TISCC feature parallel syndrome extraction with a cycle time that is constant in the code distance. The circuits are also time resolved according to literature-derived gate timings, enabling accurate resource estimation as well as the quantification of physical coherent error buildup (see Sec. III B for details). In Fig. 1, which was adapted from Ref. [21], we provide an overview of our trapped-ion surface code mapping.

While the logical actions of the output circuits of TISCC were previously verified through stabilizer simulation, verification of the fault-tolerance properties of those circuits was left to later work. In the present study, we confirm the exponential scaling (with code distance $d$) of subthreshold logical error rates for the most elementary logical operation output by TISCC: the logical idle operation. This operation consists of $d$ rounds of parallel syndrome extraction on a surface code patch encoding an arbitrary single-qubit logical state. The methodology that we use to quantify the logical error rate of this operation will be fully described in Sec. IV.

TABLE I. The physical-layer error model that we use in this work is a simplified and updated version of the trapped-ion QCCD error models described in Refs. [7,24]. Updated physical error rates have been extracted from the Quantinuum H2-1E emulator specification sheet [49]. The error rate for the measurement operation is a (generous) compromise between values from the biased measurement error model from Ref. [50]. The coherent dephasing error that accumulates during physical qubit movement and idling is a single-qubit $Z$-axis rotation modeled by $R_Z(\theta) = e^{-iZ\theta/2}$, where $\theta =$ rate $\times$ duration [24].

| Operation | Error model | Probability |
|---|---|---|
| Initialization | Bit-flip | $4.0 \times 10^{-5}$ |
| $X/Y$ single-qubit rotation | Depolarizing | $2.9 \times 10^{-5}$ |
| Two-qubit rotation | Depolarizing | $1.28 \times 10^{-3}$ |
| Measurement | Bit-flip | $1.0 \times 10^{-3}$ |
| Movement/idling | Coherent | Rate: 0.043 Hz |

### B. Physical-layer error model

In our work, we use a simplified version of the trapped-ion QCCD error model from Refs. [7,24] with more recent parameters extracted from the Quantinuum H2-1E specification sheet obtained through private communication [49], most of which is available on Github [50]. See Table I for details. It was noted in Ref. [24] that dephasing error sources resulting from, e.g., external magnetic field fluctuations, which manifest during physical qubit idling and movement, were likely to be the cause of an observed logical (Pauli) error model for the Steane code being biased toward logical $\bar{Z}$ noise. In that work, dephasing errors were modeled either as coherent errors or as the stochastic error channels that result from Pauli twirling the coherent errors. In the coherent case, the error is expressed mathematically as a $Z$-axis rotation [$R_Z(\theta) = e^{-iZ\theta/2}$] by an angle $\theta =$ rate $\times$ duration, where the duration is given by the time elapsed between gates applied to physical qubits. We refer to the rate at which coherent dephasing error accumulates as the "dephasing rate," following Ref. [24]. The Pauli twirl of this rotation (when expressed as an error channel) yields a stochastic dephasing channel with an error rate of $p_{\text{dephasing}} = \sin^2(\theta/2)$. In both Refs. [7,24], the dephasing rate was taken to be constant over all circuit locations, although this is a simplification that is not expected to be accurate [7]. Here, we use the same simplification.

We have decided to further simplify the error model we use compared with the one presented in Ref. [24] in order to focus our attention on the effects of coherent dephasing noise on logical error. As such, we do not include leakage errors or spontaneous emission (the most destructive aspects of which are also modeled by leakage) since leakage errors can be mitigated by leakage repumping [24,51]; accordingly, these errors were replaced in Quantinuum's later simulation work by depolarizing noise [7]. We have also decided to neglect crosstalk errors during physical qubit initialization and measurement because recent crosstalk error rates appear to be small [49,50,52]. Thus, we directly use gate infidelities from Quantinuum's randomized benchmarking experiments as error rates for single-qubit and two-qubit depolarizing channels following single-qubit and two-qubit unitary gates, respectively, as well as their experimental state preparation and measurement fault probabilities as error rates for bit-flip channels following (preceding) physical qubit initialization (measurement) [50]. In all of our calculations, we hold all error parameters constant at the values expressed in Table I except for the dephasing rate, which we vary.

In TISCC, several operations are implied to take place during the two-qubit gate [$R_{ZZ}(\pi/2)$] time (Split, Merge, and Cool; see Ref. [21]). The noise model we employ in the present work does not include any coherent error that may be expected to accumulate during these operations, but instead inserts coherent dephasing noise according to the time elapsed between the application of gates on physical qubits, using the time-resolved circuit schedule output by TISCC and the gate durations found in Fig. 1(c) to calculate these idling durations (we note that $Z$-axis rotations, which are assumed to be performed in software, also contribute to idling time). Thus, though we perform full circuit-level coherent noise simulations in this study, our noise model likely underestimates the amount of physical coherent noise that would be present in the TISC under the architectural assumptions of current-generation trapped-ion QCCD systems. Since it is expected that future-generation systems will pipeline operations within gate zones to minimize temporal overhead due to ion cooling [41], future work should incorporate these assumptions.

### C. Distribution of idling durations

In order to provide readers with a point of reference for considering how physical qubit movement and idling error strengths compare with the other error parameters from Table I in our circuits, we show the distribution of durations of time that elapse between physical-layer gates operating on the same qubit within TISCC idle circuits in Fig. 2. These time durations form one of the inputs to the coherent dephasing error discussed in Sec. III B alongside the dephasing rate, which we vary in our simulations. We see that the distribution of idling durations is very structured, with only a few durations having any weight. We also see that the distribution barely changes with code distance. However, as code distance increases, the relative share of large (>1 ms) durations decreases. This is because idling durations of this scale are present only at the boundaries of the surface code patch where qubits may not be in the support of four stabilizers, causing them to idle during some cnot layers. As this happens only at boundaries, its effects become less pronounced in large patches. Accordingly, the average idling durations (dashed lines) decrease with code distance. If coherent errors were to accumulate during the operations that are implied to take place during the $R_{ZZ}(\pi/2)$ gate, then the densities above 1 ms in Fig. 2 would increase accordingly.

## IV. SIMULATION METHODOLOGY

In order to emulate the TISC, we require a simulation methodology that is able to accommodate non-Clifford circuits so that we can include circuit-level coherent noise, since arbitrary-angle $Z$ rotations are not in the Clifford group and therefore cannot be simulated by standard methods such as the one from Ref. [53]. We also note that the methods of Ref. [32], in which arbitrary-angle $Z$-axis rotations are efficiently simulated for the surface code in a code capacity coherent noise model, and Ref. [54], in which the surface code under a mixed stochastic and coherent code capacity noise model is efficiently simulated, cannot be directly applied to our circuit-level simulations. To overcome these challenges, we utilize a technique involving phase-sensitive stabilizer simulation of Clifford circuits such as is done in Ref. [38] and obtain expectation values of logical observables through a Monte Carlo sampling algorithm such as is done in Ref. [37]. From these expectation values, we perform logical process tomography and extract the diamond distance from identity as our logical error rate, as we describe in Sec. IV C. We begin by discussing our Monte Carlo technique in Sec. IV A and our gate decomposition in Sec. IV B.

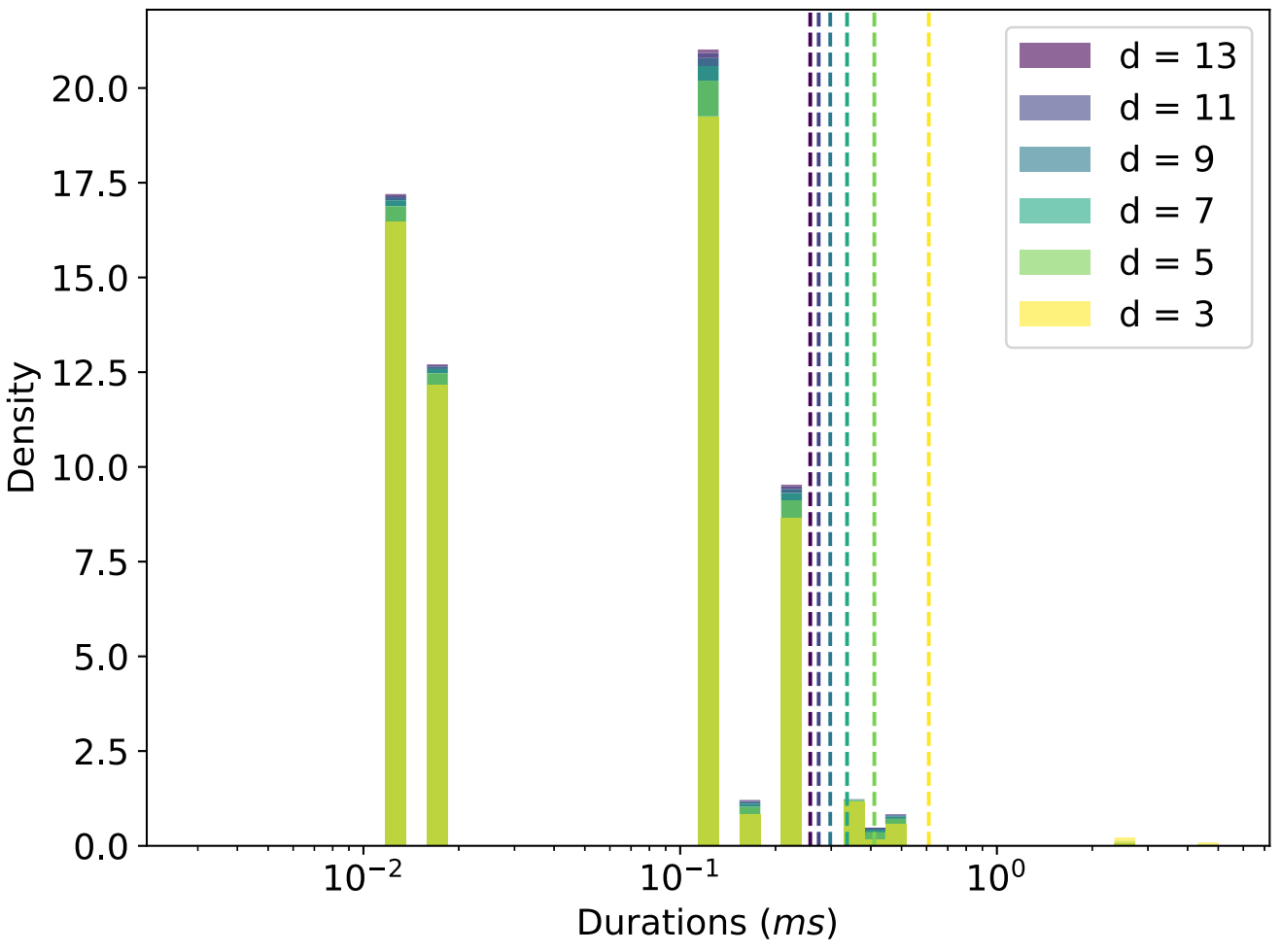

FIG. 2. Distribution of idling durations for physical qubits during a TISCC idle operation for each of the different code distances under consideration in this study. Physical qubit idling times are determined by tracking the amount of time that elapses between gates applied to physical qubits according to the time-resolved circuit schedule output by TISCC. Horizontal dashed lines represent averages.

### A. Monte Carlo sampling strategy for quasi-clifford circuits

The goal of our simulator, called the ORQCS, is to compute expectation values of the form $\mathrm{Tr}(\rho_C H)$, where $H$ is some Hermitian operator and $\rho_C$ is the result of applying a noisy near-Clifford circuit to $|0\rangle^n$. In a nutshell, our simulation strategy consists of (1) expressing circuit elements as linear combinations of Clifford operators and Pauli measurements, (2) randomly sampling and simulating Clifford sequences from these expressions, and (3) computing weighted averages of many Clifford sequences with weights derived from the expansion coefficients. This strategy is similar to that employed in previous works such as Refs. [37,38,55,56], with sample cost that grows exponentially in the "non-Cliffordness" (quantified below) of the circuit. Compared to Ref. [37], this method uses a different form of Clifford expansion that leads to better scaling with respect to the non-Cliffordness of the circuit. The two main differences of our approach to Refs. [38,55,56] are that (1) we are interested in computing expectation values of quantum observables (such as logical Pauli operators of a QECC) rather than sampling output bit strings, and (2) we allow the gates to be noisy, i.e., not unitary. This leads to incoherent contributions to expectation values. While our current implementation scales quadratically worse with non-Cliffordness than the sum-over-Cliffords method of Ref. [38], the techniques introduced in Ref. [57] could be incorporated to match the scaling of that algorithm. For the phase-sensitive stabilizer simulation, we employ the C-H representation and associated update rules described in Ref. [38].

#### 1. *Problem formulation*

The standard simulation problem is to compute expectation values of various observables at the output of a quantum circuit $C$. For the present exposition, we suppose the circuit involves no measurements; measurements will be addressed later. We assume that the input to $C$ is the state $|0\cdots0\rangle\langle0\cdots0|$. The output state $\rho_C$ may be written as

$$\rho_C = \mathcal{C}(|0\cdots0\rangle\langle0\cdots0|), \tag{1}$$

where $\mathcal{C}$ is the quantum channel representing the entire circuit. Suppose $C$ consists of $L$ (possibly imperfect) gates. Then $\mathcal{C} = \mathcal{G}_L \circ \cdots \circ \mathcal{G}_1$, where $\mathcal{G}_l$ is the quantum channel for the $l$th gate, yielding

$$\rho_C = \mathcal{G}_L \circ \cdots \circ \mathcal{G}_1(|0\cdots0\rangle\langle0\cdots0|). \tag{2}$$

Let $P^{(1)}, \ldots, P^{(K)}$ denote the observables to be measured. Without loss of generality, we take these to be multiqubit Pauli operators (any observable on a set of qubits can be written as a linear combination of such operators). The goal is to compute

$$\langle P^{(k)}\rangle = \mathrm{Tr}(\rho_C P^{(k)}) \tag{3}$$

for each $k = 1, \ldots, K$.

#### 2. *Gate decomposition*

Let $\rho_l$ denote the state after $l$ gates, with $\rho_0 = |0\rangle\langle0|^{\otimes n}$. Let $\mathcal{G}_l$ denote the quantum channel describing the effect of the $l$th gate:

$$\rho_l = \mathcal{G}_l(\rho_{l-1}). \tag{4}$$

To leverage efficient stabilizer simulation, we express $\mathcal{G}_l$ in terms of Clifford operations. To accommodate both non-Cliffordness and noise, we perform a *double decomposition* of $\mathcal{G}_l$. We first perform a Kraus decomposition of $\mathcal{G}_l$:

$$\mathcal{G}_l(\cdot) = \sum_r p_r^{(l)} K^{(l,r)}(\cdot)K^{(l,r)\dagger}, \tag{5}$$

where $p_r^{(l)}$ is a probability distribution (to uniquely define $p_r^{(l)}$, we impose the condition $\mathrm{Tr}(K^{(l,r)\dagger}K^{(l,r)}) = 2^k$, where $k$ is the number of qubits on which $K$ acts). We then decompose each $K^{(l,r)}$ as a superposition of Clifford operators:

$$K^{(l,r)} = \sum_i \alpha_i^{(l,r)} C_i. \tag{6}$$

Altogether $\rho_l$ can be expressed in terms of $\rho_{l-1}$ as

$$\rho_l = \sum_r p_r^{(l)} K^{(l,r)} \rho_{l-1} K^{(l,r)\dagger} \tag{7}$$

$$= \sum_r p_r^{(l)} \sum_{i,j} \alpha_i^{(l,r)} \alpha_j^{(l,r)*} C_i \rho_{l-1} C_j^\dagger. \tag{8}$$

There are infinitely many double decompositions of a given channel, both because Kraus decomposition is not unique and because Cliffords form a highly overcomplete operator basis. As will be evident later, it is desirable to minimize the one-norm $\|\alpha^{(l,r)}\|_1 = \sum_i |\alpha^{(l,r)}|$ of each Kraus operator expansion. The minimum one-norm that can be achieved for an operator, which we call *Clifford extent* [38], is a measure of the operator's "non-Cliffordness." For a Clifford operator, it is 1, while for non-Clifford unitaries it is greater than 1. The optimal decomposition of arbitrary channels in terms of Cliffords is an open problem. We currently perform and optimize our double decompositions manually (see Sec. IV B yfor an optimal decomposition of the Pauli rotation channel). In our experience, each one- or two-qubit gate decomposition typically consists of only a few Kraus operators and a few Cliffords per Kraus operator.

#### 3. Expansion of the circuit output state

Substituting the gate expansions (7) into Eq. (2) yields the following expansion of the output state:

$$\rho_C = \sum_{r,i,j} p_r \alpha_i^{(r)} \alpha_j^{(r)*} |\sigma_i\rangle\langle\sigma_j|, \tag{9}$$

where $r \equiv (r_1, \dots, r_L)$, $i \equiv (i_1, \dots, i_L)$, $j \equiv (j_1, \dots j_L)$, and

$$p_r \equiv p_{r_1}^{(1)} \cdots p_{r_L}^{(L)}, \tag{10}$$

$$\alpha_i^{(r)} \equiv \alpha_{i_1}^{(1,r_1)} \cdots \alpha_{i_L}^{(L,r_L)}, \tag{11}$$

$$|\sigma_i\rangle \equiv C_{i_L} \cdots C_{i_1} |0\rangle^{\otimes n}. \tag{12}$$

Note that $|\sigma_i\rangle$ is a stabilizer state obtained by applying a particular sequence of Clifford gates to $|0\rangle^{\otimes n}$.

The expectation of any operator $P^{(k)}$ can likewise be expanded as

$$\langle P^{(k)}\rangle = \mathrm{Tr}(P^{(k)}\rho_C) \tag{13}$$

$$= \sum_{r,i,j} p_r \alpha_i^{(r)} \alpha_j^{(r)*} \langle\sigma_j|P^{(k)}|\sigma_i\rangle. \tag{14}$$

#### 4. Monte Carlo approach

Obviously, for any nontrivial circuit the expansion (14) will involve far too many terms to compute them all. However, $\langle P^{(k)}\rangle$ can be efficiently estimated by sampling a relatively small random subset of the terms from these nested sums. Nominally, we estimate $\rho_C$ by

$$\tilde{\rho}_C = \frac{1}{N} \sum_{(r,i,j)\sim q}^{N} w_{r,i,j} |\sigma_i\rangle\langle\sigma_j|, \tag{15}$$

where the notation adorning the summation symbol indicates that the sum is over $N$ triples $(r, i, j)$ independently sampled from a distribution $q$ to be specified in the next subsection. Here,

$$w_{r,i,j} = \frac{p_r \alpha_i^{(r)} \alpha_j^{(r)*}}{q_{r,i,j}} \tag{16}$$

is the (complex) shot weight. Similarly, $\langle P^{(k)}\rangle$ can be estimated by

$$\tilde{P}^{(k)} = \mathrm{Tr}(P^{(k)}\tilde{\rho}_C) \tag{17}$$

$$= \frac{1}{N} \sum_{(r,i,j)\sim q}^{N} w_{r,i,j} \langle\sigma_j|P^{(k)}|\sigma_i\rangle. \tag{18}$$

It is a standard exercise to show that $\tilde{\rho}_C$ and $\tilde{P}^{(k)}$ are unbiased estimators of $\rho_C$ and $\mathrm{Tr}(P^{(k)}\rho_C)$, respectively. Note that each triple $(r, i, j)$ specifies a pair of Clifford circuits (one acting on the "ket" and one acting on the "bra") and corresponding output values $\langle\sigma_j|P^{(1)}|\sigma_i\rangle, \dots, \langle\sigma_j|P^{(K)}|\sigma_i\rangle$ in addition to the shot weight $w_{r,i,j}$. All of these quantities can be computed efficiently using a stabilizer simulator that keeps track of global phase; we use the CH-form for stabilizer states and associated simulation algorithm proposed in Ref. [38]. The simulation of a randomly sampled pair of Clifford circuits and their contribution to the final observables is called a "shot."

#### 5. Estimator variance

The variance of the estimator $\tilde{P}^{(k)}$ is

$$\mathrm{Var}[\tilde{P}^{(k)}] = \frac{1}{N}(\langle|w_{r,i,j}\langle\sigma_j|P^{(k)}|\sigma_i\rangle|^2\rangle - \langle P^{(k)}\rangle^2). \tag{19}$$

For a given set of gate decompositions, the variance depends on the sampling distribution $q$. We follow the common practice [58] of using a distribution that is a product of the distributions that are individually optimal for each gate:

$$q_{r,i,j} = \prod_{l=1}^{L} p_{r_l}^{(l)} \frac{|\alpha_{i_l}^{(l,r_l)} \alpha_{j_l}^{(l,r_l)*}|}{\|\alpha^{(l,r_l)}\|_1^2}. \tag{20}$$

This distribution is simple to implement and can be expected to be close to optimal for most observables. With this distribution, the shot weight simplifies to

$$w_{r,i,j} = \prod_{l=1}^{L} w_{r_l,i_l,j_l}, \tag{21}$$

where

$$w_{r_l,i_l,j_l} = \|\alpha^{(l,r_l)}\|_1^2 \frac{\alpha_{i_l}^{(l,r_l)} \alpha_{j_l}^{(l,r_l)*}}{|\alpha_{i_l}^{(l,r_l)} \alpha_{j_l}^{(l,r_l)*}|} \tag{22}$$

is the shot weight of gate $l$. Then,

$$\mathrm{Var}[\tilde{P}^{(k)}] = \frac{1}{N}\left(\left\langle\left|\|\alpha^{(r)}\|_1^2 \langle\sigma_j|P^{(k)}|\sigma_i\rangle\right|^2\right\rangle - \langle P^{(k)}\rangle^2\right) \tag{23}$$

$$\leqslant \frac{1}{N}\langle\|\alpha^{(r)}\|_1^4\rangle, \tag{24}$$

where $\|\alpha^{(r)}\|_1 = \prod_{l=1}^{L} \|\alpha^{(l,r_l)}\|_1$. Here, we have used the fact that $|\langle\sigma_j|P^{(k)}|\sigma_i\rangle| \leqslant 1$. This upper bound is a good proxy for the true variance in the important case that $|w_{r,i,j}| \gg 1$. In a typical scenario, each gate is a noisy unitary whose Kraus operators have an average Clifford extent $\geqslant 1$. Thus, $\|\alpha^{(r)}\|_1$ tends to grow exponentially with the number of gates and their

non-Cliffordness. If all the gates are Clifford or near-Clifford gates, one may write

$$\left\|\alpha^{(r)}\right\|_1^4 \approx \exp(\eta_1 + \cdots + \eta_L), \tag{25}$$

where $\eta_l = \langle \|\alpha^{(l,r_l)}\|_1^4 \rangle - 1$ quantifies the non-Cliffordness of the double decomposition of gate $l$. The exponent $\eta = \eta_1 + \cdots + \eta_L$ is a convenient measure of the total non-Cliffordness of the circuit.

#### 6. Simulation of mid-circuit measurements and resets

Midcircuit measurement, feedforward, and resetting of ancilla qubits are essential elements of quantum error correction. Although measurement gates fit within the previously described formalism, it is more natural to use a sampling strategy in which the probabilities of the Kraus operators (which in this case are projectors for the various measurement outcomes) are not fixed but state-dependent.

We consider measurements of Pauli operators. A noisy measurement of a Pauli operator $P$ is modeled as a noisy channel (which can be decomposed and simulated as described above) followed by a perfect measurement of $P$. We write $P = \Pi_+ - \Pi_-$, where $\Pi_\pm$ is the projector onto the $\pm 1$ eigenspace of $P$. Let $\rho = |\sigma_i\rangle\langle\sigma_j|$ denote the simulated state up to the point of measurement. Naively, one would select outcome $\pm 1$ with probability $\mathrm{Tr}(\Pi_\pm \rho)$ and set the postmeasurement state to $\Pi_\pm \rho \Pi_\pm / \sqrt{\mathrm{Tr}(\Pi_\pm \rho)}$. But in our simulation approach $\rho$ is not necessarily a density operator, which means that the traces in question may vanish and the corresponding probabilities and postmeasurement states may not be well defined. Instead, we first compute the four probabilities:

$$q_i^\pm = \langle \sigma_i | \Pi_\pm | \sigma_i \rangle, \tag{26}$$

$$q_j^\pm = \langle \sigma_j | \Pi_\pm | \sigma_j \rangle. \tag{27}$$

Then, the projections of the input state onto the eigenspaces of $P$ can be written as

$$\Pi_\pm |\sigma_i\rangle\langle\sigma_j| \Pi_\pm = w p_\pm |\sigma_i^\pm\rangle\langle\sigma_j^\pm|, \tag{28}$$

where

$$|\sigma_i^\pm\rangle = \frac{\Pi_\pm|\sigma_i\rangle}{\sqrt{q_i^\pm}}, \quad |\sigma_j^\pm\rangle = \frac{\Pi_\pm|\sigma_j\rangle}{\sqrt{q_j^\pm}} \tag{29}$$

are normalized stabilizer states,

$$p_\pm = \frac{\sqrt{q_i^\pm q_j^\pm}}{\sqrt{q_i^+ q_j^+} + \sqrt{q_i^- q_j^-}} \tag{30}$$

are proper probabilities that sum to 1, and

$$w = \sqrt{q_i^+ q_j^+} + \sqrt{q_i^- q_j^-} \tag{31}$$

is the shot weight factor for this gate. Simulation is performed by selecting outcome $\pm 1$ with probability $p_\pm$, computing the corresponding postmeasurement states $|\sigma_i^\pm\rangle$ and $|\sigma_j^\pm\rangle$, and multiplying the shot weight by $w$. Simulation then proceeds with the next gate.

A related operation is "reset $P$" which deterministically sets the eigenvalue of $P$ to $+1$. Regardless of how it is physically implemented, it can be modeled as an implicit measurement of $P$ followed by a conditional operation that flips the eigenvalue to $+1$ if the $-1$ eigenvalue was observed. Let $Q$ be a Pauli that acts on the same qubits as $P$ and anticommutes with $P$. The reset-$P$ operation is implemented by performing a measurement of $P$ as described above, then in the case of the $-1$ outcome, applying $Q$ to both $|\sigma_i^-\rangle$ and $|\sigma_j^-\rangle$.

In contrast to (noisy) unitary channels, the shot weight factor $w$ of a measure or reset gate can be less than 1. This can help mitigate the typically exponential growth of the circuit shot weight for non-Clifford circuits. It is even possible to have $w = 0$; in this case, the shot can be aborted, since the final shot weight will be zero and the shot will contribute nothing to any estimator.

#### 7. Cost analysis

The total cost of the algorithm is the cost per shot times the number of shots needed to obtain a desired variance $\epsilon^2$. On each shot, two output stabilizer states and $K$ expectation values are calculated. The cost of simulating one local Clifford gate ranges from $O(n)$ to $O(n^3)$ depending on the type of gate, so the cost of computing $|\sigma_i^{(r)}\rangle$ and $|\sigma_j^{(r)}\rangle$ is $O(Ln^3)$. To compute $\langle\sigma_j^{(r)}|P^{(k)}|\sigma_i^{(r)}\rangle$, we apply $P^{(k)}$ to $|\sigma_i^{(r)}\rangle$ to obtain a stabilizer state and compute the inner product of two stabilizer states, which has cost $O(n^3)$. Thus, the total cost for one shot is $O((2L+K)n^3)$. To obtain a variance $\epsilon^2$, the number of shots must be

$$N = \frac{1}{\epsilon^2}\left\langle \left\|\alpha^{(r)}\right\|_1^4 \right\rangle. \tag{32}$$

It follows that the total cost of this approach is

$$\mathrm{cost} = \frac{1}{\epsilon^2}\left\langle \left\|\alpha^{(r)}\right\|_1^4 \right\rangle O((2L+K)n^3). \tag{33}$$

The most expensive part of our current implementation is the calculation of inner products of stabilizer states $\langle\sigma_j|\sigma_i\rangle$, which cost $O(n^3)$. We believe that this can be reduced to $O(n^2)$ via an alternative stabilizer representation. It is also worth noting that the technique employed in Ref. [57] would yield quadratically better scaling with respect to the Clifford extent $\|\alpha^{(r)}\|_1$, but calculating observables would require computation of a much larger number of inner products.

### B. Decomposition of dephasing error

The circuits in this study consist entirely of Clifford gates. The error models for initialization, one- and two-qubit gates, and measurement can all be represented as probabilistic mixtures of Clifford gates. Thus, the only non-Clifford aspect of the simulation is the dephasing error, represented by $Z(\theta) \equiv \exp(-\mathrm{i}\frac{\theta}{2}Z)$, where $\theta$ is the accumulated phase. The corresponding channel

$$\mathbf{Z}(\theta)(\rho) = Z(\theta)\rho Z(\theta)^\dagger \tag{34}$$

is already in Kraus form. It remains to obtain a Clifford decomposition of $Z(\theta)$. For $0 \leqslant \theta \leqslant \pi/4$, the optimal Clifford decomposition of $Z(\theta)$ is [38]

$$Z(\theta) = \left(\cos\frac{\theta}{2} - \sin\frac{\theta}{2}\right)I + \left(\sqrt{-2i}\sin\frac{\theta}{2}\right)S, \qquad (35)$$

where $S$ is the phase gate. Optimal decompositions for other values of $\theta$ are similar and can be obtained by symmetry arguments, e.g., subtract off the nearest multiple of $\pi/2$ and insert a corresponding number of factors of $S$.

### C. Computational workflow

The simulation method described in the previous section is one part of our computational workflow for calculating logical error rates for the mapping described in Sec. III. Briefly, TISCC is used to generate time-resolved hardware circuits for the surface code operation of interest at a given code distance (in this study, we focus only on the idling surface code) [21]. In the next stage of the pipeline, TISCC circuits are read into ORQCS, which inserts appropriate noise operations into the circuit. Quantum channel representations for each noise source except coherent dephasing noise are built at this stage. The coherent dephasing channel, on the other hand, is built on-the-fly during simulation according to physical qubit idling times, which are tracked by the simulator. For each shot $s$, two Clifford circuits are independently simulated in a phase-sensitive fashion [38], and inner products $\langle\sigma_{j_s}|P_L|\sigma_{k_s}\rangle$ are calculated between the final stabilizer states $|\sigma_{j_s}\rangle$ and $P_L|\sigma_{k_s}\rangle$, where $P_L \in \{X_L, Y_L, Z_L\}$ are the logical Pauli observables of the surface code patch. Each shot also yields an associated shot weight $w_{r_s,k_s,j_s}$ [see Eq. (16)] and set of measurement outcomes $\{m_i\}_s$, where the index $i$ labels the sequence of measurements in the TISCC circuit. These measurement outcomes are used to decode the aforementioned inner products using minimum-weight perfect matching (MWPM). Decoded results are combined with sample weights to estimate the expectation values of $P_L$ according to Eq. (18). In what follows, we provide further detail about certain steps of this process, including how we extract logical error rates from our estimated expectation values.

#### *1. Decoding*

We consider two different decoding graph constructions, both of which rely on PyMatching [59,60]. First, in a code path that we call “PCM” (for parity check matrix), we directly use the surface code parity check matrix output by TISCC with PyMatching to construct a matching graph with equal weights for all error mechanisms (including both spacelike and timelike edges). In a second code path that we call “DEM” (for detector error model), we use Stim [53] to produce a so-called detector error model for circuits that can be used in subsequent matching graph construction. The detector error model is a general way to express which deterministic combinations of measurement outcomes, or detectors, are flipped by circuit-level errors and with what probability. The DEM code path involves translating TISCC circuits into Stim circuits and adding in our error model but with Pauli-twirled coherent errors. Conveniently, PyMatching is able to construct a matching graph directly from Stim’s DetectorErrorModel object.

Our preliminary results demonstrated that the DEM and PCM code paths produced qualitatively similar results, although the DEM code path yielded much lower logical error rates. This is to be expected, since the decoding graph constructed in the DEM code path uses edge weights that more accurately reflect the error model of Sec. III B than the equal weights of the PCM code path. Since, for the purposes of this study, we are most interested in the qualitative features of our results, and since the higher logical error rates in the PCM code path enabled better statistics with fewer samples, all of the results presented in Sec. V use the PCM code path. For each sample, we form detectors from the measurement outcomes and decode them using MWPM on one of the matching graph constructions (either PCM or DEM). The result of PCM decoding is a physical-layer correction; we check whether this correction anticommutes with any of the logical Pauli observables $P_L$ and update them if so. The result of DEM decoding is a logical-layer correction: The decoder determines whether $X_L$ and/or $Z_L$ flipped, and the XOR of these results determines whether $Y_L$ flipped. Then, with $\{\pm\langle\sigma_{j_s}|P_L|\sigma_{k_s}\rangle\}$ and $w_{r_s,k_s,j_s}$, we may use Eq. (18) to obtain estimates of $\{\langle P_L\rangle\}$. It is worth noting at this point that our expectation values are not conditioned on syndrome measurements, but instead represent averages over all samples. This will have implications on the physical interpretation of the logical process matrices we derive from these expectation values, as we discuss later.

#### *2. Logical process tomography*

These outputs enable TISCC and ORQCS to be used together within a logical process tomography workflow. Using TISCC, we obtain logical state preparation circuits to prepare states $\{|\alpha\rangle_L\}$ from the set $\{|Z_L, +1\rangle, |Z_L, -1\rangle, |X_L, +1\rangle, |Y_L, +1\rangle\}$, which we relabel (in order of appearance) as $\{|0\rangle_L, |1\rangle_L, |+\rangle_L, |-\rangle_L\}$. The outputs of these logical state preparation circuits were validated in the noiseless case using logical state tomography in Ref. [21]. To simulate one shot of the noisy logical idle operation, we include noiseless state preparation, $d$ rounds of noisy syndrome extraction, and a single perfect round of syndrome extraction (to ensure that final stabilizer states are projected into a logical subspace, enabling single-qubit process tomography [61]) and do not perform a logical measurement before calculating $\langle\sigma_{j_s}|P_L|\sigma_{k_s}\rangle$. Upon obtaining estimates of $\{\langle P_L\rangle\}$ for each logical input state, we use either of two techniques to calculate process matrices from them. First, we use the standard linear inversion technique from Ref. [62] to calculate the Chi matrix. Alternatively, we may straightforwardly obtain the PTM elements from our expectation values as follows.

If $\mathcal{N}(\rho)$ is a single-qubit quantum process acting on state $\rho$, it can be expressed in terms of its action on elements of the Pauli basis as

$$\mathcal{N}(\rho) = \sum_{\sigma} |\mathcal{N}(\sigma)\rangle\rangle\langle\langle\sigma|\rho\rangle\rangle, \qquad (36)$$

where $\sigma \in \{I, X, Y, Z\}/\sqrt{2}$ are the normalized Pauli matrices and the notation $|\rho\rangle\rangle$ denotes a vectorized quantum state with elements $\langle\langle\sigma|\rho\rangle\rangle \equiv \mathrm{Tr}[\sigma\rho]$. Since $|\mathcal{N}(\sigma)\rangle\rangle = \sum_\sigma \mathrm{Tr}[\sigma\mathcal{N}(\sigma)]|\sigma\rangle\rangle$, Eq. (36) can be rewritten as

$$\mathcal{N}(\rho) = \sum_\sigma \sum_{\sigma'} \mathrm{Tr}[\sigma'\mathcal{N}(\sigma)]|\sigma'\rangle\rangle\langle\langle\sigma|\rho\rangle\rangle. \tag{37}$$

From here, we may define the PTM representation of $\mathcal{N}(\rho)$ as a matrix whose elements are given by $\mathcal{N}_{\sigma',\sigma} \equiv \mathrm{Tr}[\sigma'\mathcal{N}(\sigma)]$. Since $|\rho\rangle\rangle$ is a vector containing the trace of $\rho$ (which should always be equal to one) as well as a Bloch vector of its Pauli expectation values, Eq. (37) shows that the PTM encodes the redistribution of Pauli expectation values across the Pauli basis as well as the preservation of trace and the unitality of the process. Loosely, for example, $\mathcal{N}_{\sigma_1,\sigma_2}$ tells us how much of the original $\langle Y\rangle$ gets transferred to $\langle X\rangle$. Further, the leftmost column, $\mathcal{N}_{\sigma',\sigma_0}$, quantifies the unitality of $\mathcal{N}(\rho)$, while its topmost row, $\mathcal{N}_{\sigma_0,\sigma}$, quantifies trace preservation. Note that, for logical qubits, we take $\sigma_L \in \{I_L, X_L, Y_L, Z_L\}/\sqrt{2}$ to be the normalized logical Pauli operators $P_L/\sqrt{2}$. Although one might wish to define the logical subspace with respect to a single stabilizer subspace by including a projector in the definition of $\sigma_L$ (most commonly $\Pi_0$, the projector onto the $+1$ eigenspace of all stabilizers [28]), this results in an unrealistic scenario for fault-tolerant state preparation, as we discuss later.

To calculate $\mathcal{N}_{\sigma',\sigma}$ from our estimates of $\{\langle P_L\rangle\}$ from each logical input state, we may first write (dropping subscripts $L$ for brevity of notation) the Pauli matrices in terms of our input states:

$$\sigma_0 = (|0\rangle\langle 0| + |1\rangle\langle 1|)/\sqrt{2}, \tag{38}$$

$$\sigma_1 = (2|+\rangle\langle +| - |0\rangle\langle 0| - |1\rangle\langle 1|)/\sqrt{2}, \tag{39}$$

$$\sigma_2 = (2|-\rangle\langle -| - |0\rangle\langle 0| - |1\rangle\langle 1|)/\sqrt{2}, \tag{40}$$

$$\sigma_3 = (|0\rangle\langle 0| - |1\rangle\langle 1|)/\sqrt{2}. \tag{41}$$

From here, we see that the columns of the PTM can be written as (defining $\langle\sigma\rangle_{|\alpha\rangle} \equiv \mathrm{Tr}[\sigma\mathcal{N}(|\alpha\rangle\langle\alpha|)]$)

$$\mathcal{N}_{\sigma',\sigma_0} = (\langle\sigma'\rangle_{|0\rangle} + \langle\sigma'\rangle_{|1\rangle})/\sqrt{2}, \tag{42}$$

$$\mathcal{N}_{\sigma',\sigma_1} = (2\langle\sigma'\rangle_{|+\rangle} - \langle\sigma'\rangle_{|0\rangle} - \langle\sigma'\rangle_{|1\rangle})/\sqrt{2}, \tag{43}$$

$$\mathcal{N}_{\sigma',\sigma_2} = (2\langle\sigma'\rangle_{|-\rangle} - \langle\sigma'\rangle_{|0\rangle} - \langle\sigma'\rangle_{|1\rangle})/\sqrt{2}, \tag{44}$$

$$\mathcal{N}_{\sigma',\sigma_3} = (\langle\sigma'\rangle_{|0\rangle} - \langle\sigma'\rangle_{|1\rangle})/\sqrt{2}. \tag{45}$$

When we use our estimates for $\{\langle P_L\rangle_{|\alpha\rangle}\}$, factors of $\sqrt{2}$ combine to yield a denominator of 2 in these expressions. As a comment, if the channel is unital, i.e., $\mathcal{N}_{\sigma',\sigma_0} = 0$ for $\sigma' \in \{\sigma_1, \sigma_2, \sigma_3\}$, then there is a one-to-one correspondence between the elements of $\mathcal{N}_{\sigma',\sigma}$ and the estimates $\{\langle P_L\rangle_{|\alpha\rangle}\}$ output by our simulation workflow.

The logical process tomography method that we have described in this section results in a syndrome-averaged logical process matrix. Error rates that we compute from these averaged channels are distinct from those that are calculated from syndrome-conditioned channels and subsequently averaged together, as demonstrated in Ref. [32]. Since coherent logical error results from constructive interference between errors projected onto the same syndrome sector, it is expected to be suppressed when contributions from many sectors are stochastically added together [29]. It has, however, been suggested that the averaged channel will provide the most practical characterization for large-scale fault-tolerant quantum computing [29]. In practice, we do not expect our simulation method to enable us to scalably compute separate process matrices for each syndrome sector.

Also, while our logical state preparation circuits are noiseless, we allow for random stabilizer initialization instead of initializing all stabilizers to the $+1$ eigenspace using, e.g., a unitary encoding circuit. Since unitary encoding circuits are usually not fault tolerant (see, e.g., Ref. [63]), the standard initialization protocol for surface codes is to simply project the physical qubits (initialized in a product state) onto a random stabilizer subspace (also known as a quiescent state) by performing syndrome extraction $d$ times [10,64]. Interestingly, it has been found that the impact of physical coherent noise depends on syndrome sector, with the $+1$ eigenspace causing the most constructive interference and thus being the most vulnerable to this noise [65]. Thus, allowing random stabilizer initialization is not only simpler but also expected to suppress coherent logical errors.

### 3. *Diamond error*

The diamond error provides a rigorous notion of logical error rate that is commonly used in studies involving coherent noise due to its connection with threshold theorems and its relationship to the total variation distance between actual and ideal output distributions following quantum circuits [29,32,66–69]. It can be thought of as the quantum channel analog to the trace distance between quantum states. For the noisy logical idle operation under consideration in this study, $\mathcal{N}(\rho)$, the diamond error is its diamond distance from the logical identity process $D_\diamond \equiv ||\mathcal{N}(\rho) - \mathcal{I}||_\diamond$. We note that this definition of logical error rate has a maximum of 2. Other commonly used measures of error (such as average infidelity) cannot be used to estimate the diamond error for processes involving coherent noise [67]. One can calculate the diamond error given logical process matrices the ideal and actual quantum processes using semidefinite programs [70]. In this work, we use the Qiskit implementation [71].

## V. RESULTS

In this section, we present our results. First, we provide the logical Pauli expectation value estimates we obtain from an informationally complete set of input logical states (see Sec. IV C), including their sample variances. We discuss their relevance, make connections to the PTM representation of the logical error channel where possible, and make observations regarding the limitations of our simulator along the way. Finally, we present logical error rates with some caution, avoiding combinations of code distance and dephasing rate for which the error PTM elements had large variances.

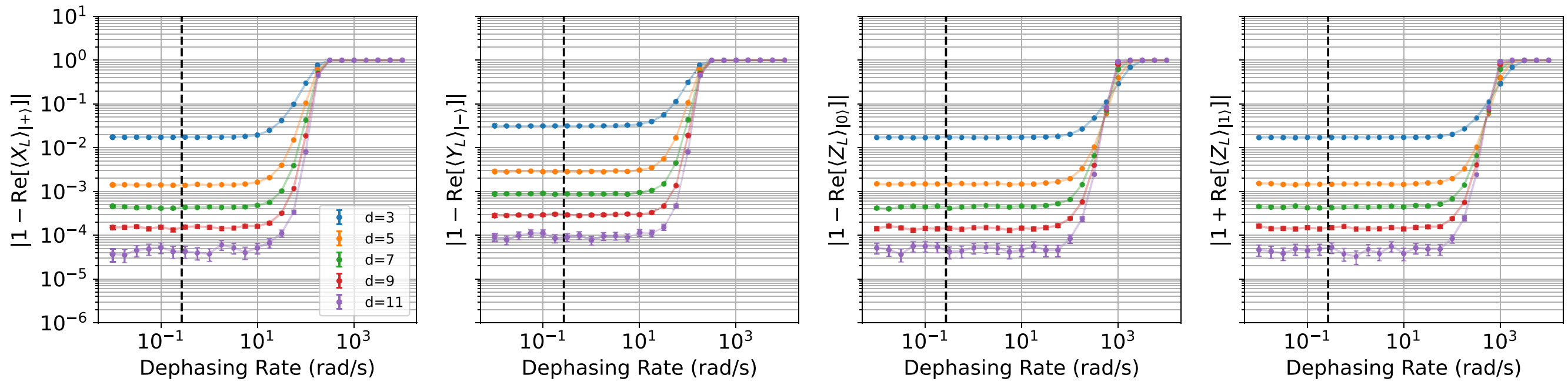

FIG. 3. Diagonal expectation value estimates for our fully stochastic simulations at various code distances and dephasing rates. Error bars are given by $2\epsilon$, where $\epsilon$ is the standard deviation of estimator $\tilde{P}_{L,|\alpha\rangle}$ for the logical Pauli operator in question. The dashed black line corresponds with the dephasing rate of current-generation Quantinuum hardware.

### A. Expectation values and the Pauli transfer matrix

First, we report our expectation value estimates $\langle P_L\rangle_{|\alpha\rangle} \equiv \mathrm{Tr}[P_L\mathcal{N}(|\alpha\rangle\langle\alpha|)]$ for all combinations of logical input states $\{|\alpha\rangle\}$ and logical Pauli observables $P_L$ under consideration (see Sec. IV C for details). Since, in general, our expectation value estimates can have small imaginary components, which are unphysical and independent of the real components, we report only the real parts of these estimates. More specifically, where $\mathrm{Tr}[P_L|\alpha\rangle\langle\alpha|] = \pm 1$, we report estimates of $|1 \mp \mathrm{Re}[\langle P_L\rangle_{|\alpha\rangle}]|$, and where $\mathrm{Tr}[P_L|\alpha\rangle\langle\alpha|] = 0$, we report estimates of $|\mathrm{Re}[\langle P_L\rangle_{|\alpha\rangle}]|$, in order to inspect quantities related to the PTM of the logical error process $\mathcal{E} \equiv |\mathcal{N}(\rho) - \mathcal{I}|$. We refer to these two kinds of expectation values as "diagonal" and "off-diagonal" expectation values, respectively, due to their relationships with the diagonal and off-diagonal elements of the logical error PTM [see Eqs. (42)–(45)].

In what follows, we compare results obtained from simulations under the noise model described in Sec. III B with those obtained from simulations under the equivalent noise model where the coherent dephasing channel is replaced by its Pauli twirl. We call these respective versions of the noise model the "mixed coherent and stochastic" and "fully stochastic" noise models. In both cases, we hold all of the parameters of the noise model constant except the dephasing rate, which we vary. For our simulations of the fully stochastic noise model, we ran a modest number of samples, $5 \times 10^6$ at $d = 3 - 9$ and $2 \times 10^6$ at $d = 11$, which we found sufficient to deliver results with reasonably small error bars and to enable us to understand qualitative trends. The mixed coherent and stochastic simulations, on the other hand, required many more samples to obtain results with acceptable error bars, which we liberally define as $2\epsilon \leqslant x$, where $\epsilon$ is the standard deviation of estimate $x$. Given the scaling relationship from Eq. (32), we expect that the number of samples necessary to obtain a given variance increases both with the code distance and with the coherent dephasing rate, and we have confirmed this expectation in practice. For instance, our $d = 3$–$5$ results include between $5 \times 10^6$ samples at the lowest dephasing rates and $4 \times 10^9$ at the highest dephasing rates. The number of samples we collected at $d = 7 - 11$ lies between these two extremes. The standard deviations of our expectation value estimates, scaled by the number of samples taken, are provided in Appendix A.

#### *1. Fully stochastic noise model*

In Fig. 3, we show expectation value estimates obtained from simulations under the fully stochastic noise model. Since it is impossible to have coherent logical rotations under this noise model, we only show the expectation values of logical Paulis for which the input state is an eigenstate (the other combinations of observable and input state yield expectation values that are exactly zero). First, we notice from the plots that thresholds lie well above the published dephasing rate of current-generation Quantinuum systems, which is marked with a dashed line. This means that our TISC mapping effectively minimizes qubit movement and idling time to the extent that the associated dephasing noise does not pose an issue in this regime (at least as far as these stochastic simulations are concerned). Second, we notice that the dephasing rate thresholds are similar between the plots of $\langle X_L\rangle_{|+\rangle}$ and $\langle Y_L\rangle_{|-\rangle}$, but that these plots suggest different thresholds from the plots of $\langle Z_L\rangle_{|0\rangle}$ and $\langle Z_L\rangle_{|1\rangle}$. We suspect that this is because, while the syndrome extraction circuit will convert $R_Z(\theta)$ errors into $R_X(\theta)$ or $R_Y(\theta)$ errors in some fault locations (particularly those pertaining to qubit movement in the midst of cnot gate execution), the strongest dephasing errors (which correspond to the longest idling times) occur while boundary data qubits are idling between cnot layers (see Sec. III C), and these do not have converted axes. Since only rotations about Pauli axes that anticommute with a given logical operator can cause errors for that operator, we find different dephasing rate thresholds for different logicals, with that of $Z_L$ being higher than those of $X_L$ and $Y_L$. Third, while the error process in question is expected to be unital due to the lack of, e.g., amplitude damping processes in the physical-layer noise model we simulate, it is clear that, because of the probabilistic nature of our simulations, $\langle Z_L\rangle_{|0\rangle}$ and $\langle Z_L\rangle_{|1\rangle}$ will not exactly cancel out. Assuming unitality, however, the first three panels of Fig. 3 contain the only nontrivial elements of the logical error process' PTM [see Eqs. (42)–(45)].

#### *2. Mixed coherent and stochastic noise model*

In Fig. 4, we present the analogous (diagonal) results obtained from simulations under the mixed coherent and stochastic noise model. Since our simulation algorithm is not expected to scale well for large dephasing rates due to the resulting circuits being more highly non-Clifford, we probe lower dephasing rates in Fig. 4 than we do in Fig. 3. As anticipated, the range of dephasing rates for which we were able to obtain accurate results differed by code distance: While we ran simulations for dephasing rates from $10^{-6}$ rad/s through $10^{2}$ rad/s for $d = 3$–$5$, we only considered values from $10^{-6}$ rad/s through $10^{1}$ rad/s for $d = 7$–$11$, and not all of the grid points we probed yielded acceptable error bars. As such, in Fig. 4 we fade data points for which the half-widths of the error bars, here given by two standard deviations $\epsilon$, are more than 100% of the value of the expectation value estimate itself. The results in Fig. 4 are consistent with those of Fig. 3 away from the threshold region. Interestingly, however, the results in Fig. 4 are suggestive of a reduced threshold for the TISC under the mixed coherent and stochastic noise model as compared with the TISC under the fully stochastic noise model. If one considers only unfaded points to draw this conclusion, it is evident from the plots of $|1 - \mathrm{Re}[\langle Z_L\rangle_{|0\rangle}]|$ at $d = 3$–$5$ (third panel of Fig. 4) that the data series cross before 100 rad/s. This conclusion is corroborated by the analogous $|1 + \mathrm{Re}[\langle Z_L\rangle_{|1\rangle}]|$ data in panel four, although the point at 100 rad/s is faded in the $d = 5$ plot due to a relatively large variance. Further, threshold reductions for the $X_L$ and $Y_L$ logicals are suggested by the plots of $|1 - \mathrm{Re}[\langle X_L\rangle_{|+\rangle}]|$ and $|1 - \mathrm{Re}[\langle Y_L\rangle_{|-\rangle}]|$ at $d = 3$–$5$ (first and second panels of Fig. 4, respectively), where faded $d = 5$ data points at 100 rad/s very nearly cross the $d = 3$ data series. Despite these observations, it is difficult to make confident claims based on these data due to the large variances present in our results at higher dephasing rates and code distances.

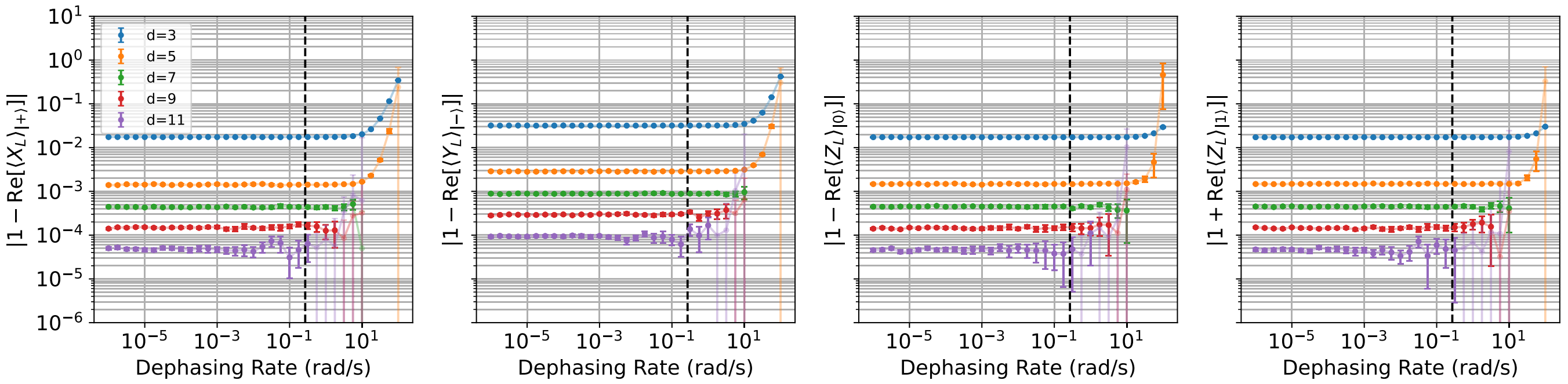

FIG. 4. Diagonal expectation value estimates for our mixed coherent and stochastic simulations at various code distances and dephasing rates. Error bars are given by $2\epsilon$, where $\epsilon$ is the standard deviation of estimator $\mathrm{Re}[\tilde{P}_{L,|\alpha\rangle}]$ for the Pauli expectation value in question. Where $|1 \mp \mathrm{Re}[\tilde{P}_{L,|\alpha\rangle}]|$ is plotted, unfaded data satisfy the condition $2\epsilon \leqslant |1 \mp \mathrm{Re}[\tilde{P}_{L,|\alpha\rangle}]|$. The dashed black line corresponds with the dephasing rate of current-generation Quantinuum hardware.

As in the fully stochastic simulations, $\langle Z_L\rangle_{|0\rangle}$ and $\langle Z_L\rangle_{|1\rangle}$ will not exactly cancel out, causing spurious deviations from unitality (a similar observation holds for other $\langle P_L\rangle_{|0\rangle}$ and $\langle P_L\rangle_{|1\rangle}$ where nonzero). Barring this, as in Fig. 3, the first three panels of Fig. 4 contain the nontrivial diagonal elements of the corresponding logical error process' PTM. Therefore, if the trends observed in the preceding paragraph for these diagonal expectation values at $d = 3$–$5$ were to hold, they would confirm that coherent physical noise manifests in a logical error PTM that has larger diagonal elements near the threshold than its fully stochastic counterpart (this is especially evident from the plots of $|1 - \mathrm{Re}[\langle Z_L\rangle_{|0\rangle}]|$ at $d = 5$). Larger diagonal elements in the presence of physical coherent noise were suggested by Ref. [28] as underlying the common observation that subthreshold logical error rates are larger under coherent physical noise than under stochastic physical noise. Our findings suggest that this is true only very near to the threshold in our circuit-level noise model. Indeed, for $d \geqslant 7$, we see no significant difference between the results in Figs. 3 and 4. Finally, we comment that our results suggest a threshold for $Z_L$ that is lower than that of $X_L$ or $Y_L$ under our mixed coherent and stochastic noise model, which defies our expectations based on the comments made about noise conversion in the discussion of Fig. 3.

Next, we turn our attention to off-diagonal expectation values under the mixed coherent and stochastic noise model. Where nonzero, these expectation values provide evidence for coherent logical noise. To emphasize this, in Fig. 5 we organize our results by the logical axis that a corresponding logical rotation would be about. In particular, Figs. 5(a)–5(c) show expectation value estimates which, under the assumption of unitality, correspond to off-diagonal elements of the PTM associated with coherent logical rotations about the $X_L$, $Y_L$, and $Z_L$ axes, respectively. As in Fig. 4, we fade out data points for which the half-width error bars are more than 100% of the corresponding expectation value estimate. As we find only minimal evidence for coherent logical rotations at $d = 7$ and no evidence for them at $d = 9$–$11$, we plot results for $d = 3$–$5$ only. Especially given that we are considering averages over many syndrome sectors, it is not surprising that coherent logical error is strongly suppressed at higher code distances [29]. That said, we remind the reader that we did not run simulations for $d = 7$–$11$ above a dephasing rate of 10 rad/s, and therefore are not in a position to comment on the potential presence of coherent logical rotations in that regime. In these plots, we have set our resolution (the lower bound on the $y$ axis) to $10^{-10}$, although the number of samples we collected is variable by data point, as previously discussed.

While most of these data have quite a high variance, conclusions can still be drawn from them. First, and most significantly, we do detect the presence of coherent logical errors about all three logical Pauli axes in our simulations. Additionally, all coherent logical errors appear to be negligible at dephasing rates below the dashed black line, indicating a lack of coherent logical error for the TISC at current experimental dephasing rates. As expected, the most clearly pronounced

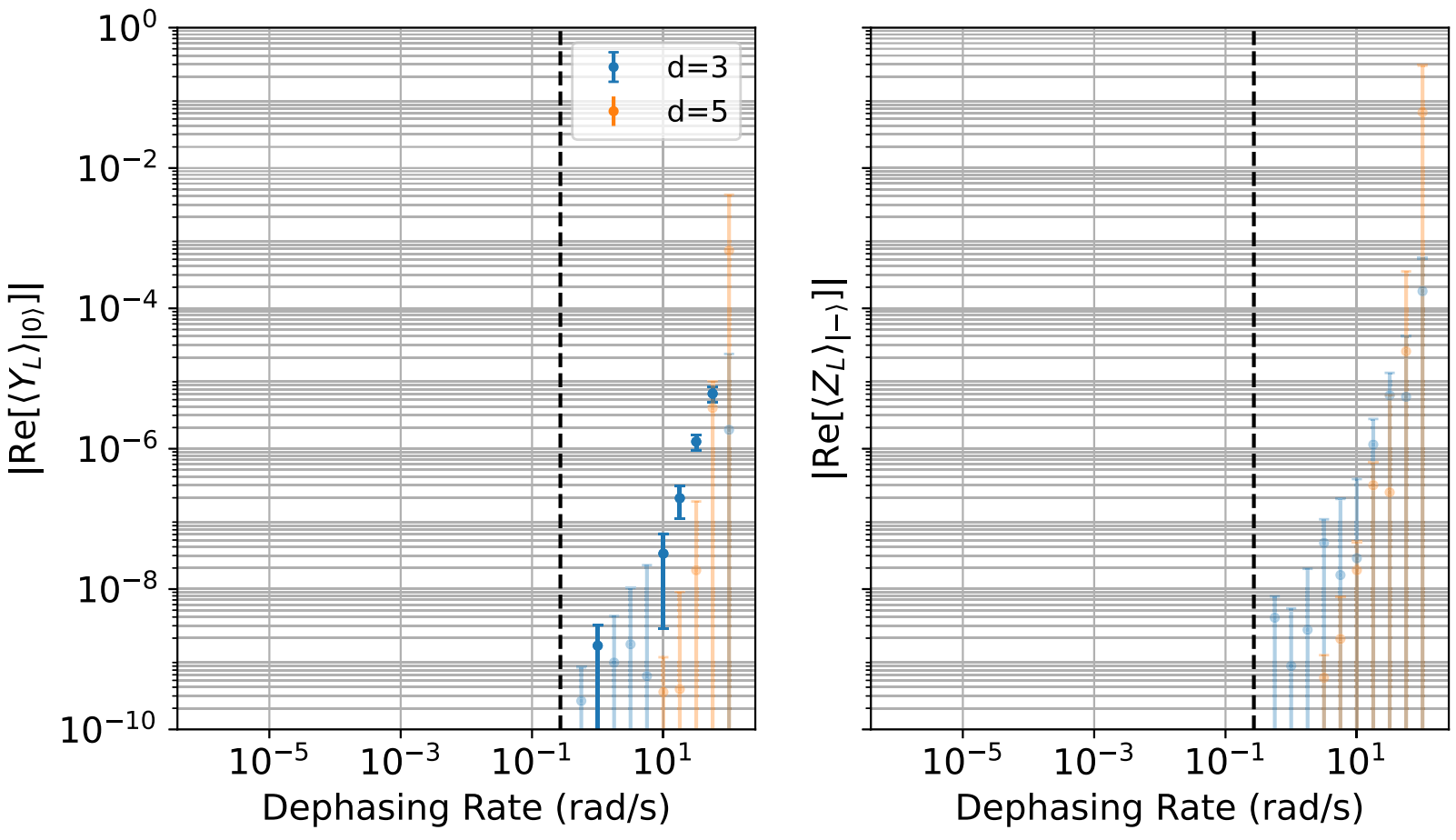

(a) Off-diagonal expectation values corresponding to a logical $X_L$ rotation.

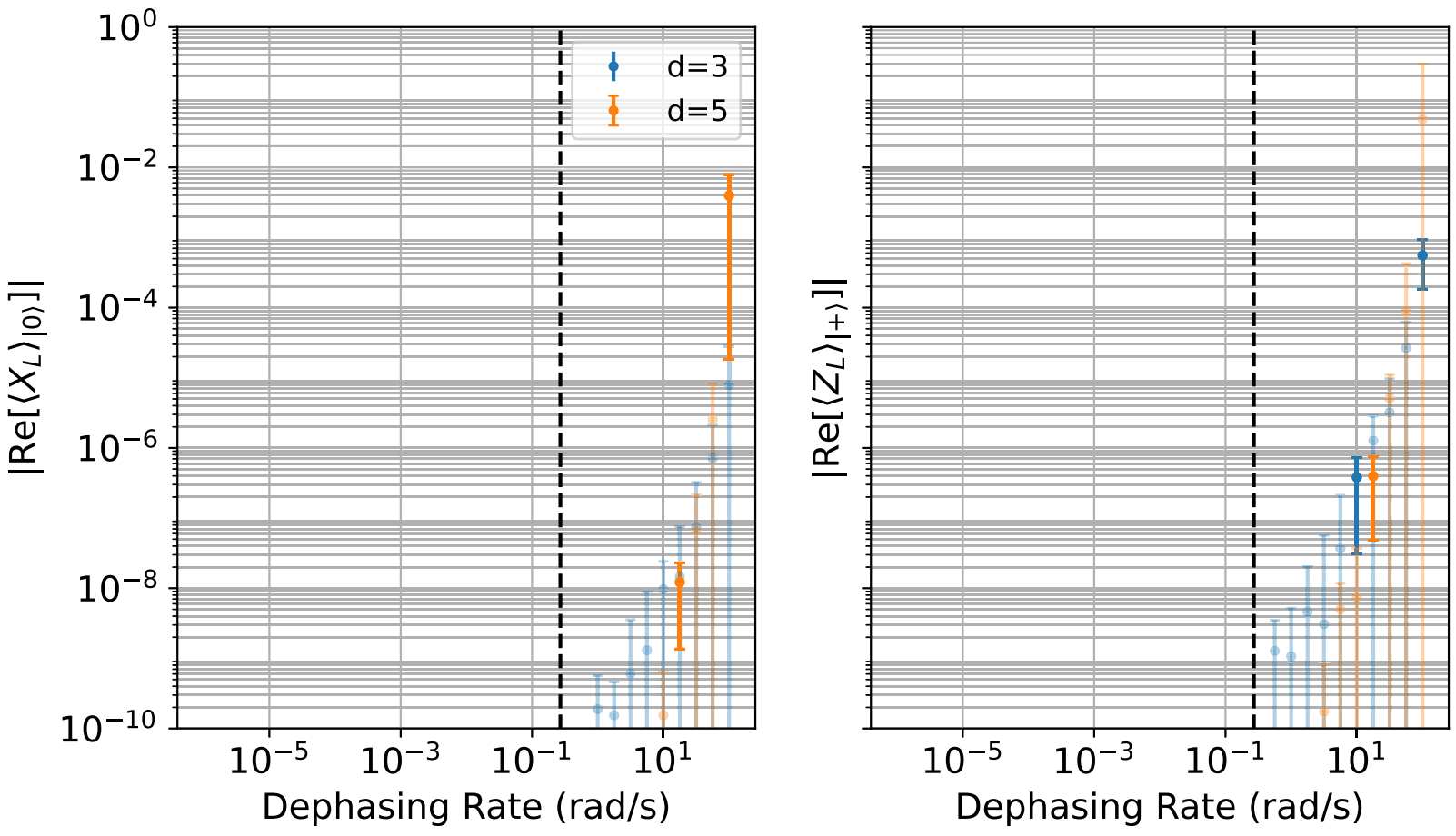

(b) Off-diagonal expectation values corresponding to a logical $Y_L$ rotation.

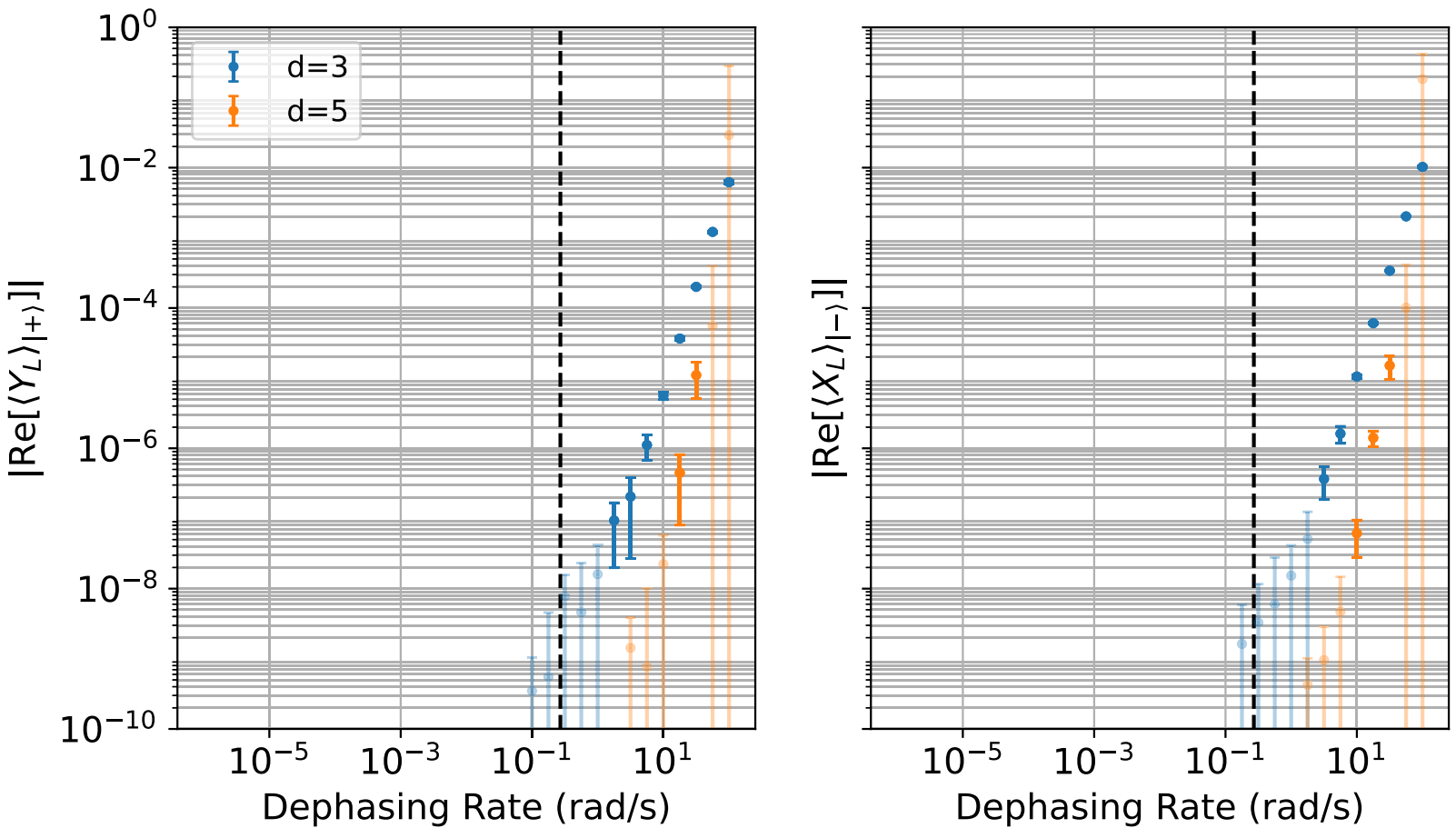

(c) Off-diagonal expectation values corresponding to a logical $Z_L$ rotation.

FIG. 5. (a)–(c) Off-diagonal expectation values corresponding with logical $X_L$, $Y_L$, and $Z_L$ rotations in our mixed coherent and stochastic simulations at various code distances and dephasing rates. Error bars are given by $2\epsilon$, where $\epsilon$ is the standard deviation of estimator $\mathrm{Re}[\tilde{P}_{L,|\alpha\rangle}]$ for the relevant Pauli expectation value. Unfaded data satisfy the condition $2\epsilon \leqslant |\mathrm{Re}[\tilde{P}_{L,|\alpha\rangle}]|$. The dashed black line corresponds with the dephasing rate of current-generation Quantinuum hardware.

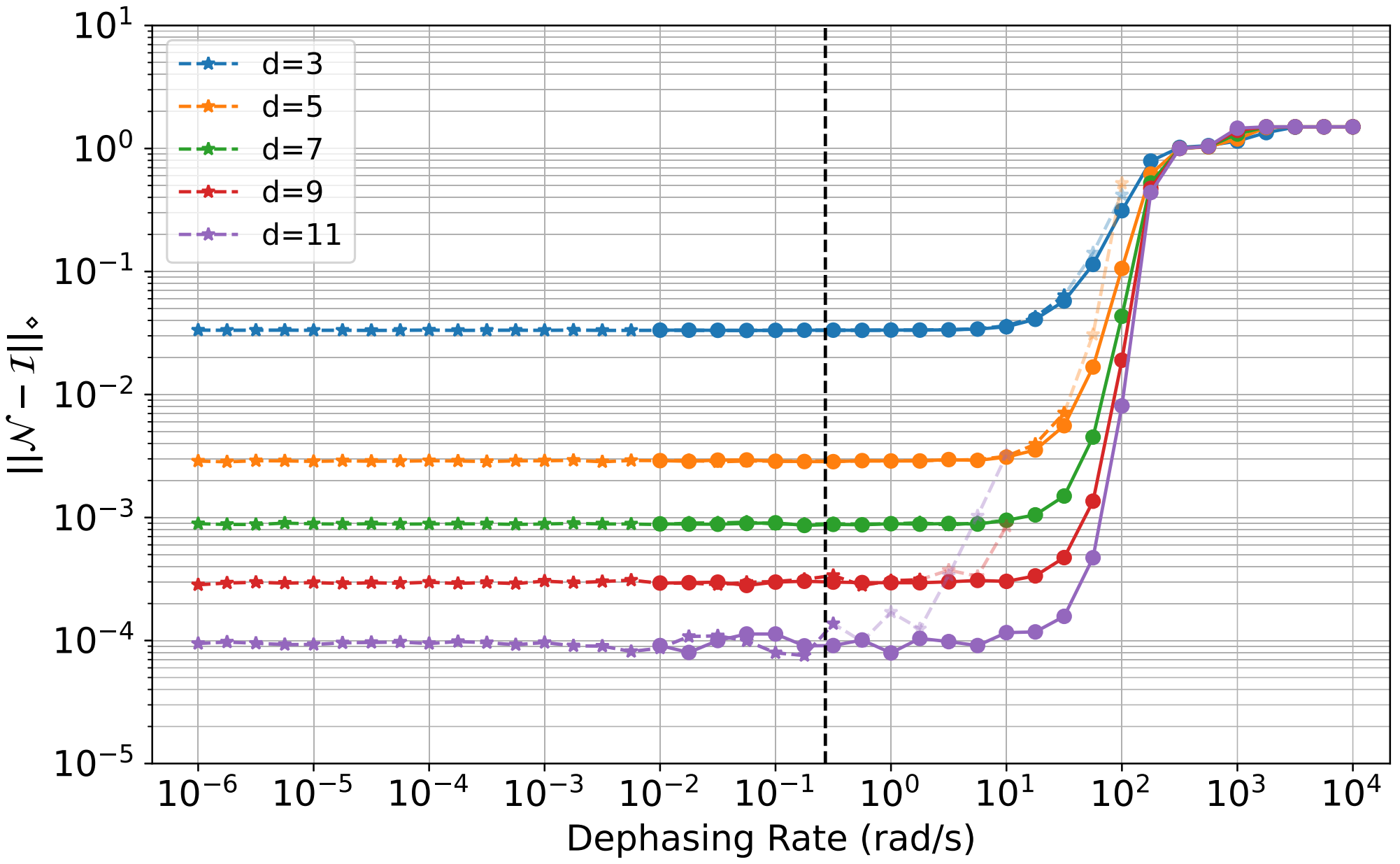

FIG. 6. Diamond error for the noisy idle operation in our mixed coherent and stochastic simulations (stars) and our fully stochastic simulations (circles) at various code distances and dephasing rates. Unfaded data are derived from logical error processes whose Pauli transfer matrix elements satisfy either the condition $2\epsilon_{\sigma',\sigma} \leqslant |\mathcal{N}_{\sigma',\sigma} - \mathcal{I}_{\sigma',\sigma}|$, where $\epsilon_{\sigma',\sigma}$ is the standard deviation of $\mathcal{N}_{\sigma',\sigma}$, or have error magnitude $|\mathcal{N}_{\sigma',\sigma} - \mathcal{I}_{\sigma',\sigma}| \leqslant 10^{-5}$. The dashed black line corresponds with the dephasing rate of current-generation Quantinuum hardware.

logical rotation is about the $Z_L$ axis [Fig. 5(c)]. Second, there appears to be an input-state-dependent asymmetry in the logical error channel. In particular, the two expectation values shown in Fig. 5(c) are asymmetric, with $|\text{Re}[\langle X_L \rangle_{|-\rangle}]|$ being a little larger than $|\text{Re}[\langle Y_L \rangle_{|+\rangle}]|$. While we have not investigated the reason for this in detail, we suspect that it is due to the greater ability of physical coherent errors to constructively interfere in $|-\rangle_L$ than in $|+\rangle_L$, because of the latter preparation circuit involving the complete randomization of $Z$ syndromes, which causes some coherent phase angles to effectively change sign. This kind of effect is known to mitigate the formation of logical coherent errors from physical coherent errors [65].

The variances of the data in Figs. 4 and 5, scaled by the number of shots, can be found in Figs. 7 and 8, respectively. In Fig. 7, it can be seen that the variance of diagonal expectation values is dominated by the stochastic part of the noise model at low dephasing rates, becomes a stable power law at intermediate dephasing rates, and begins to increase rapidly at high dephasing rates. In Fig. 8, it can be seen that the variance of off-diagonal expectation values increases very rapidly from its outset. These observations demonstrate that it would be challenging to use our simulator to probe higher dephasing rates than those we have presented in this study.

### B. Diamond error

For our calculation of logical error rates from the expectation value estimates of Sec. V A, we follow the following procedure. First, we force logical error channels to be unital by setting $\langle P_L \rangle'_{|0\rangle} = (\langle P_L \rangle_{|0\rangle} - \langle P_L \rangle_{|1\rangle})/2$, $\langle P_L \rangle'_{|1\rangle} = -\langle P_L \rangle'_{|0\rangle}$, and $\epsilon'_{P_L,|0\rangle} = \epsilon'_{P_L,|1\rangle} = (\epsilon_{P_L,|0\rangle} + \epsilon_{P_L,|1\rangle})/2$ and calculate the diamond error from the resulting channels (see Sec. IV C). This is justified on the basis that our physical-layer noise model does not include any nonunital components (see Sec. III B), and so any deviation from unitality in the logical error model

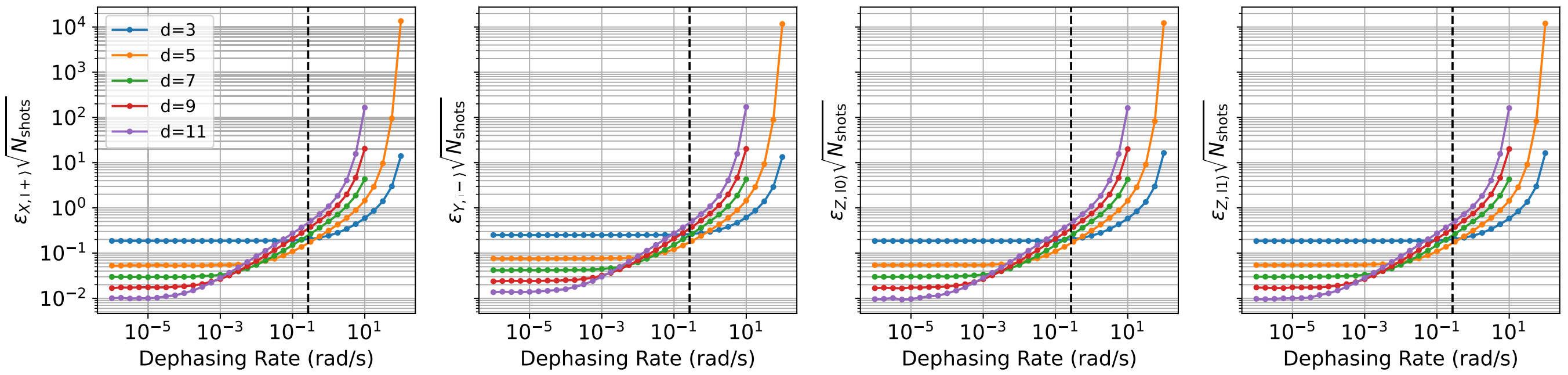

FIG. 7. One standard deviation of our estimator for each diagonal expectation value in our mixed coherent and stochastic simulations at various code distances and dephasing rates, scaled by $\sqrt{N_{\text{shots}}}$. In the $y$-axis labels, $\epsilon_{P_L,|\alpha\rangle}$ is the standard deviation for our estimate of $\langle P_L \rangle_{|\alpha\rangle}$. The dashed black line corresponds with the dephasing rate of current-generation Quantinuum hardware.

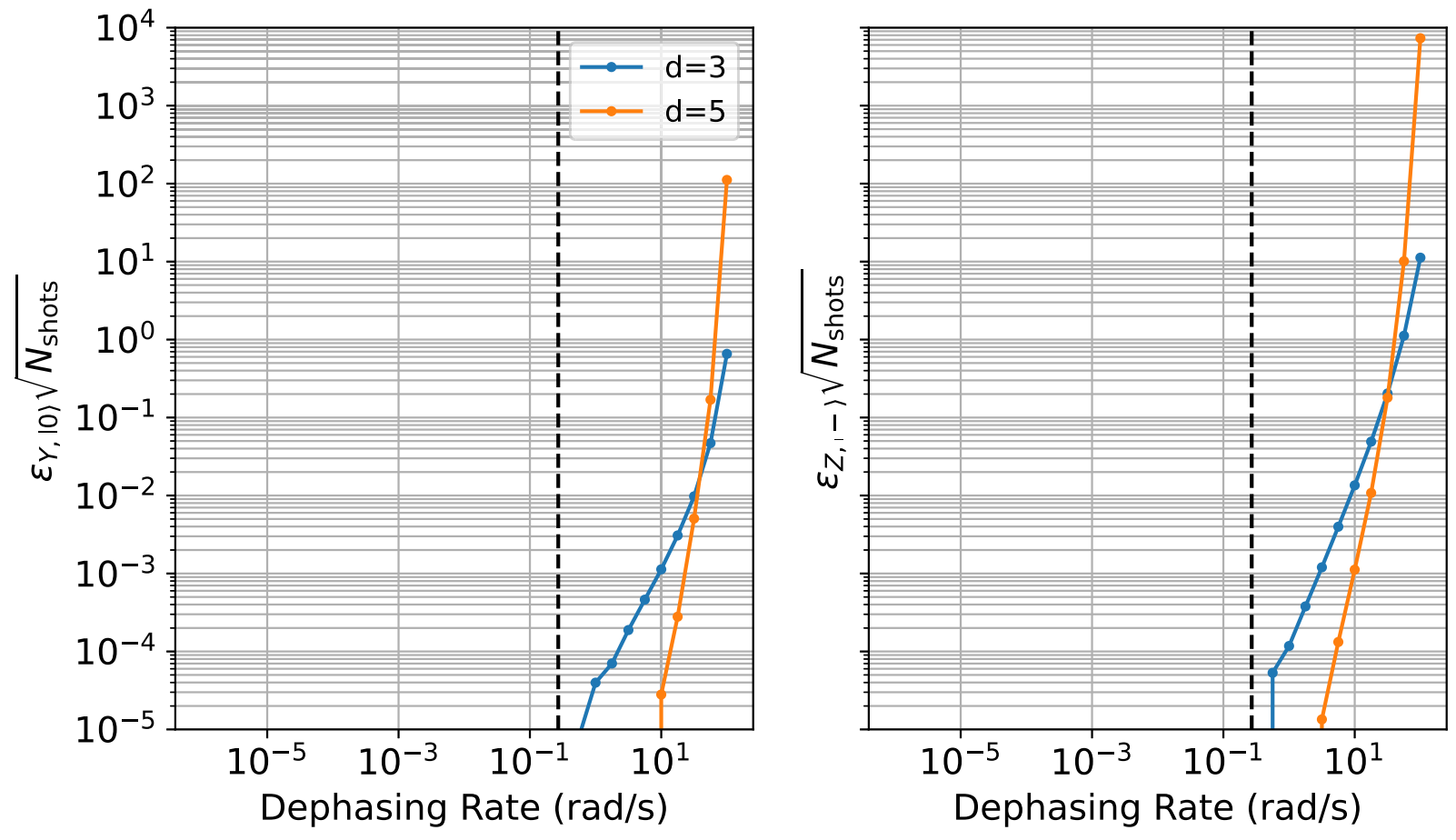

(a) Scaled standard deviation of off-diagonal expectation values corresponding to a logical $X_L$ rotation.

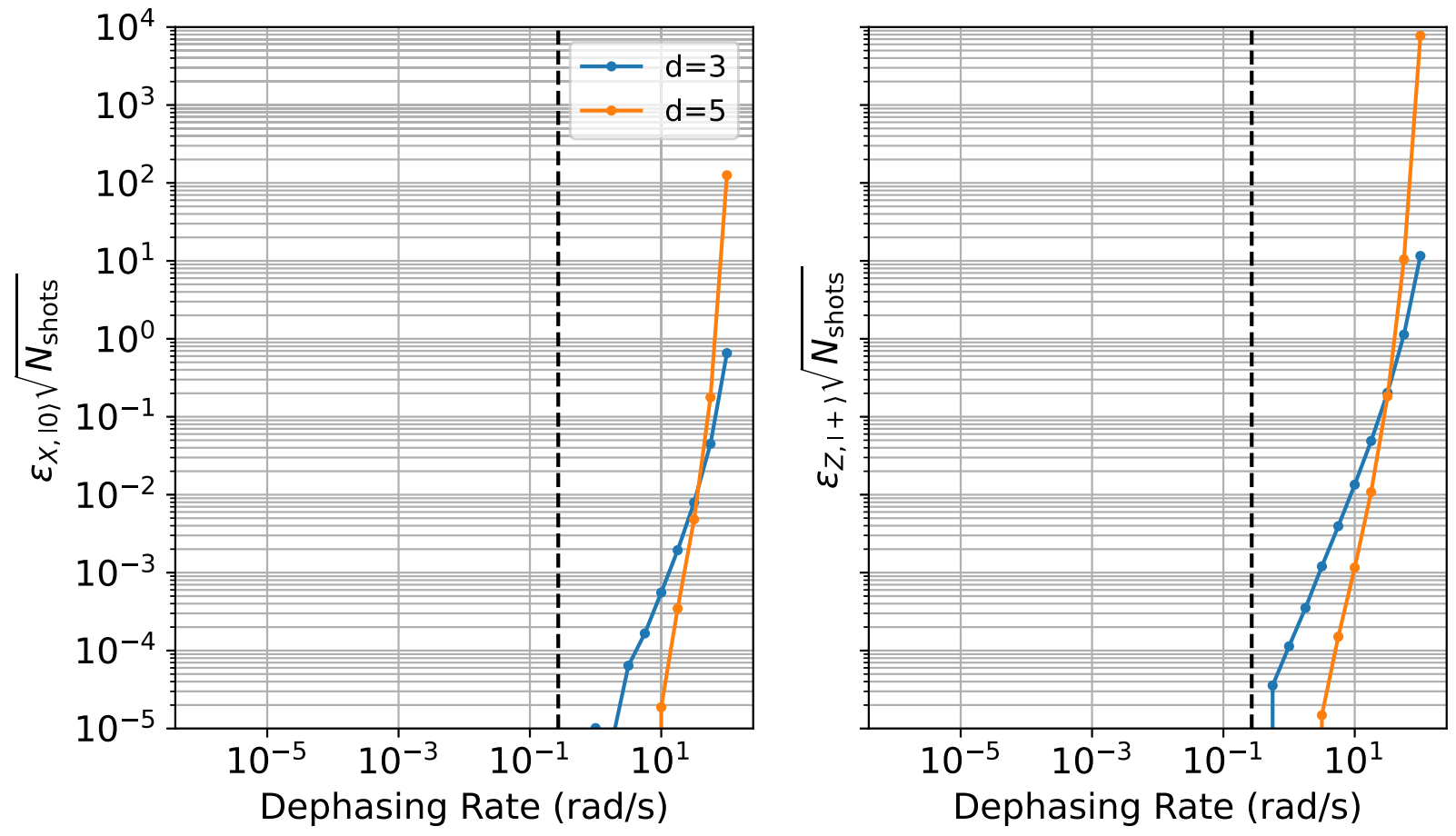

(b) Scaled standard deviation of off-diagonal expectation values corresponding to a logical $Y_L$ rotation.

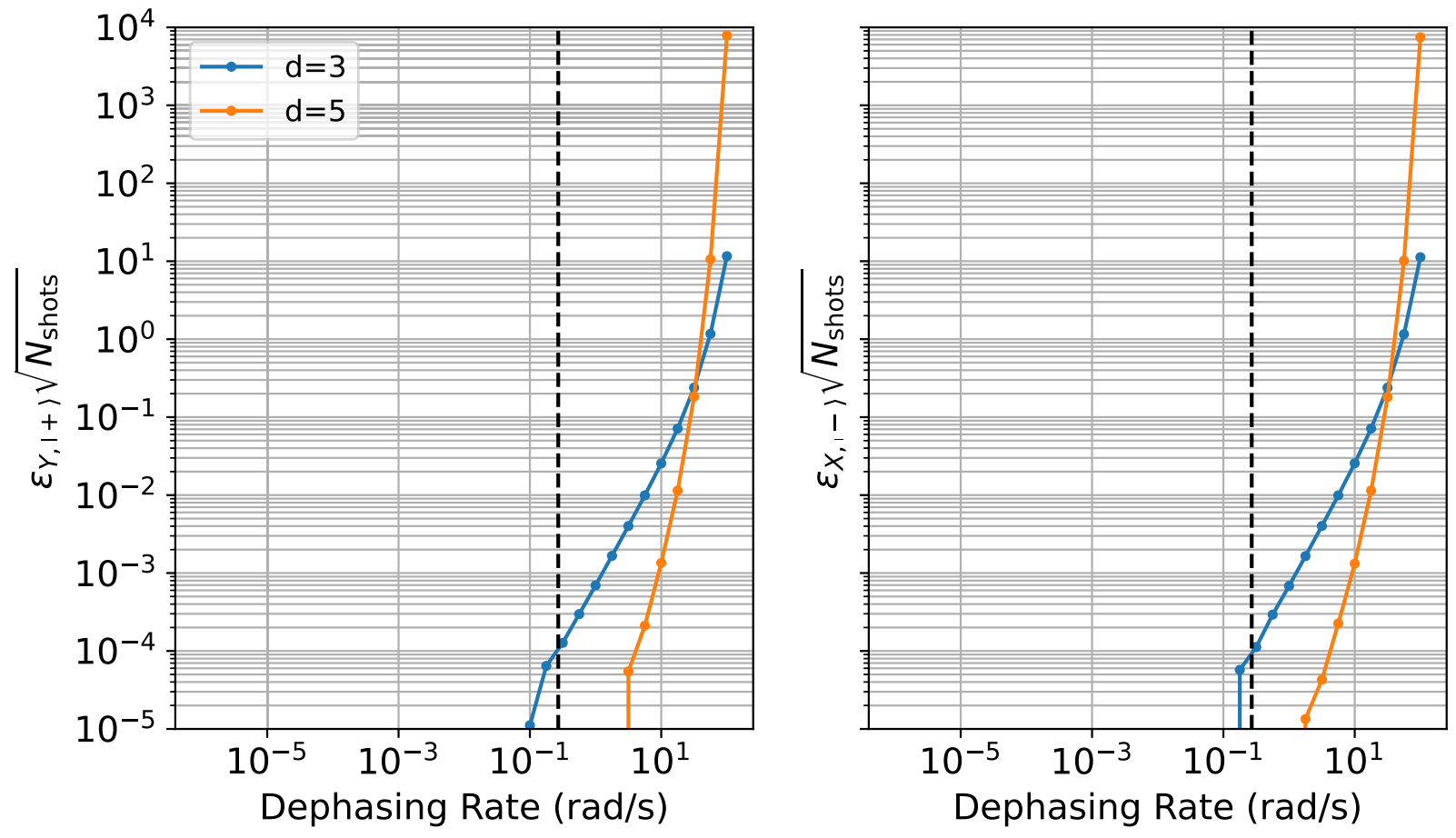

(c) Scaled standard deviation of off-diagonal expectation values corresponding to a logical $Z_L$ rotation.

FIG. 8. (a)–(c) One standard deviation of our estimator for each off-diagonal expectation value corresponding with logical $X_L$, $Y_L$, and $Z_L$ rotations in our mixed coherent and stochastic simulations at various code distances and dephasing rates, scaled by $\sqrt{N_{\text{shots}}}$. In the $y$-axis labels, $\epsilon_{P,|\alpha\rangle}$ is the standard deviation for our estimate of $\langle P_L\rangle_{|\alpha\rangle}$. The dashed black line corresponds with the dephasing rate of current-generation Quantinuum hardware.

results from the unphysical statistical fluctuations remarked upon in the discussions of Figs. 3 and 4. We deem the diamond error at a particular combination of code distance and dephasing rate to be acceptable if every logical error PTM element, $|\mathcal{N}_{\sigma',\sigma} - \mathcal{I}_{\sigma',\sigma}|$, has either an acceptable variance (according to the condition previously noted), or a magnitude $|\mathcal{N}_{\sigma',\sigma} - \mathcal{I}_{\sigma',\sigma}| \leqslant 10^{-5}$. The latter condition enables us to calculate logical error rates at grid points for which the off-diagonal expectation values are small but growing; these small values do not have a significant impact on the calculation.

Our results can be found in Fig. 6, where the fully stochastic error rates are given by filled circles with solid lines and the mixed coherent and stochastic error rates are given by stars with dashed lines. It is evident that, for the combinations of code distance and dephasing rate whose process matrices were deemed acceptable, the fully stochastic and mixed coherent and stochastic data are consistent with each other. Some evidence for a reduced threshold in our mixed coherent and stochastic simulations is developing in the $d = 3–5$ data around 100 rad/s, as discussed at length in Sec. V A. We have observed that the $d = 5$ expectation value estimates that were faded in Fig. 4 are relatively close to satisfying the acceptability criterion as compared with the faded $d = 9–11$ data in the same figure. Additionally, we have found that some expectation value estimates in the faded $d = 11$ data are not bounded by $[-1, 1]$ as expected for Pauli observables. Therefore, we have more reason to believe that the faded $d = 3–5$ data are suggestive of real qualitative threshold trends than the faded $d = 9–11$ data are. For all $d$, the diamond error approaches that of the completely depolarizing channel above ≈1000 rad/s.

## VI. CONCLUSIONS

In this work, we have presented a detailed study of logical errors in the TISC under a mixed coherent and stochastic circuit-level noise model inspired by recent trapped-ion QCCD experiments. To the best of our knowledge, this is the first study of logical error rates in such a setup above $d = 3$. In order to study circuit-level coherent noise, we have implemented a simulation methodology that relies on the Monte Carlo sampling and phase-sensitive stabilizer simulation of Clifford circuits from the Clifford decomposition of near-Clifford circuits, building upon the work of Refs. [37, 38]. Using this technique, we were able to include the coherent dephasing noise that accumulates during ion transport and idling in trapped-ion QCCD architectures [39,40] within our simulations. We probed dephasing rates near and beyond those of current-generation trapped-ion QCCD systems. Our results have the caveat that we did not allow coherent noise to accumulate during ion cooling, which we did not model explicitly.

At $d = 3–5$, we found evidence of coherent logical error about all three Pauli axes, an increased magnitude of logical error in Pauli transfer matrix diagonals near the threshold, and suggestions of a reduced threshold relative to our simulations of the analogous fully stochastic noise model. At $d = 7–11$, where we did not probe as near to the threshold region, we found close alignment between the mixed coherent and stochastic results and their fully stochastic counterparts, and no evidence of coherent logical noise (except very minimally at $d = 7$). Our results suggest that the coherence of physical qubit idling noise has a negligible impact on logical error in this regime. Although we focused our study on a mapping of the surface code to the hypothetical trapped-ion QCCD architecture proposed in Ref. [21], we note that similar considerations could apply to architectures based upon neutral atoms, which also rely upon qubit movement and suffer from physical coherent errors [3].

It would be difficult for our simulator to probe higher dephasing rates than we presented in this work. This is made clear from Figs. 7 and 8 and through the fact that, at $d = 3–5$ for the highest dephasing rates we probed, we needed to collect up to four billion samples to obtain expectation value estimates that were still quite noisy in many cases. Due to the importance of understanding the impact of physical coherent noise in realistic implementations of QECC, however, future work should improve the scaling of near-Clifford simulation methods. In parallel, future work should model logical errors for full hardware-mapped computational instruction sets; this timely task will aid us in better understanding error propagation in emerging fault-tolerant quantum computing architectures and in more accurately estimating resource requirements for executing practical quantum circuits with them.

## ACKNOWLEDGMENTS

The authors would like to acknowledge useful discussions with Balint Pato, Qile Su, and Katie Chang about modeling logical coherent noise, and Ciaran Ryan-Anderson about the Quantinuum H2-1E emulator noise model.

This research was sponsored by the Laboratory Directed Research and Development Program of Oak Ridge National Laboratory, managed by UT-Battelle, LLC, for the U.S. Department of Energy. This research used resources of the Compute and Data Environment for Science (CADES) at the Oak Ridge National Laboratory, which is supported by the Office of Science of the U.S. Department of Energy under Contract No. DE-AC05-00OR22725.

This manuscript has been authored by UT-Battelle, LLC, under Contract No. DE-AC05-00OR22725 with the U.S. Department of Energy (DOE). The US government retains and the publisher, by accepting the article for publication, acknowledges that the US government retains a nonexclusive, paidup, irrevocable, worldwide license to publish or reproduce the published form of this manuscript, or allow others to do so, for US government purposes. DOE will provide public access to these results of federally sponsored research in accordance with the DOE Public Access Plan [72].

## DATA AVAILABILITY

The data that support the findings of this article are not publicly available upon publication because it is not technically feasible and/or the cost of preparing, depositing, and hosting the data would be prohibitive within the terms of this research project. The data are available from the authors upon reasonable request.

## APPENDIX: VARIANCES

In Figs. 7 and 8, we show the standard deviations of our estimates for $\langle P_L\rangle_{|\alpha\rangle} \equiv \mathrm{Tr}[P_L\mathcal{N}(|\alpha\rangle\langle\alpha|)]$, $\epsilon_{P_L,|\alpha\rangle}$, where the estimator in question was defined in Eq. (18). The $\epsilon_{P_L,|\alpha\rangle}$ in Figs. 7 and 8 corresponds directly to the expectation value estimates in Figs. 4 and 5. Since we collected different numbers of samples for different combinations of code distance $d$ and dephasing rate (as explained in Sec. V A), we scale $\epsilon_{P_L,|\alpha\rangle}$ by $\sqrt{N_{\mathrm{shots}}}$ to obtain smooth curves. In Fig. 7, we find a regime (low dephasing rates) where stochastic noise dominates our variances followed by a regime in which these variances become dominated by coherent noise. It can be readily observed that $\epsilon_{P_L,|\alpha\rangle}$ experiences rapid growth and becomes unwieldy around $\gtrsim 10^0$ rad/s at the largest $d$ we consider. Similarly, Fig. 8 demonstrates rapid growth in the same regime for off-diagonal expectation values at $d = 3$–$5$.

[1] R. Acharya, D. A. Abanin, L. Aghababaie-Beni, I. Aleiner, T. I. Andersen, M. Ansmann, F. Arute, K. Arya, A. Asfaw, N. Astrakhantsev *et al.*, Quantum error correction below the surface code threshold, Nature (London) **638**, 920 (2024).

[2] Google Quantum AI, Suppressing quantum errors by scaling a surface code logical qubit, Nature (London) **614**, 676 (2023).

[3] D. Bluvstein, A. A. Geim, S. H. Li, S. J. Evered, J. Ataides, G. Baranes, A. Gu, T. Manovitz, M. Xu, M. Kalinowski *et al.*, Architectural mechanisms of a universal fault-tolerant quantum computer, arXiv:2506.20661.

[4] D. Bluvstein, S. J. Evered, A. A. Geim, S. H. Li, H. Zhou, T. Manovitz, S. Ebadi, M. Cain, M. Kalinowski, D. Hangleiter *et al.*, Logical quantum processor based on reconfigurable atom arrays, Nature (London) **626**, 58 (2024).

[5] A. Paetznick, M. da Silva, C. Ryan-Anderson, J. Bello-Rivas, J. Campora III, A. Chernoguzov, J. Dreiling, C. Foltz, F. Frachon, J. Gaebler *et al.*, Demonstration of logical qubits and repeated error correction with better-than-physical error rates, arXiv:2404.02280.

[6] C. Ryan-Anderson, N. Brown, C. Baldwin, J. Dreiling, C. Foltz, J. Gaebler, T. Gatterman, N. Hewitt, C. Holliman, C. Horst *et al.*, High-fidelity teleportation of a logical qubit using transversal gates and lattice surgery, Science **385**, 1327 (2024).

[7] C. Ryan-Anderson, N. Brown, M. Allman, B. Arkin, G. Asa-Attuah, C. Baldwin, J. Berg, J. Bohnet, S. Braxton, N. Burdick *et al.*, Implementing fault-tolerant entangling gates on the five-qubit code and the color code, arXiv:2208.01863.

[8] P. Sales Rodriguez, J. M. Robinson, P. N. Jepsen, Z. He, C. Duckering, C. Zhao, K.-H. Wu, J. Campo, K. Bagnall, M. Kwon *et al.*, *Experimental Demonstration of Logical Magic State Distillation* (Nature Publishing Group, London, UK, 2025), pp. 1–3.

[9] S. Dasu, S. Burton, K. Mayer, D. Amaro, J. A. Gerber, K. Gilmore, D. Gresh, D. DelVento, A. C. Potter, and D. Hayes, Breaking even with magic: Demonstration of a high-fidelity logical non-Clifford gate, arXiv:2506.14688.

[10] A. G. Fowler, M. Mariantoni, J. M. Martinis, and A. N. Cleland, Surface codes: Towards practical large-scale quantum computation, Phys. Rev. A **86**, 032324 (2012).

[11] D. Litinski, A game of surface codes: Large-scale quantum computing with lattice surgery, Quantum **3**, 128 (2019).

[12] D. Litinski and N. Nickerson, Active volume: An architecture for efficient fault-tolerant quantum computers with limited non-local connections, arXiv:2211.15465.

[13] S. Bravyi, A. W. Cross, J. M. Gambetta, D. Maslov, P. Rall, and T. J. Yoder, High-threshold and low-overhead fault-tolerant quantum memory, Nature (London) **627**, 778 (2024).

[14] Q. Xu, J. P. Bonilla Ataides, C. A. Pattison, N. Raveendran, D. Bluvstein, J. Wurtz, B. Vasić, M. D. Lukin, L. Jiang, and H. Zhou, Constant-overhead fault-tolerant quantum computation with reconfigurable atom arrays, Nat. Phys. **20**, 1084 (2024).

[15] H. Yamasaki and M. Koashi, Time-efficient constant-space-overhead fault-tolerant quantum computation, Nat. Phys. **20**, 247 (2024).

[16] H. Goto, High-performance fault-tolerant quantum computing with many-hypercube codes, Sci. Adv. **10**, eadp6388 (2024).

[17] Y. Zhao, Y. Ye, H.-L. Huang, Y. Zhang, D. Wu, H. Guan, Q. Zhu, Z. Wei, T. He, S. Cao *et al.*, Realization of an error-correcting surface code with superconducting qubits, Phys. Rev. Lett. **129**, 030501 (2022).

[18] L. Grans-Samuelsson, R. V. Mishmash, D. Aasen, C. Knapp, B. Bauer, B. Lackey, M. P. da Silva, and P. Bonderson, Improved pairwise measurement-based surface code, Quantum **8**, 1429 (2024).

[19] K. Yin, X. Fang, Z. Chen, A. Li, D. Hayes, E. Kaur, R. Nejabati, H. Haeffner, W. Campbell, E. Hudson *et al.*, Flexion: Adaptive *in-situ* encoding for on-demand QEC in ion trap systems, arXiv:2504.16303.

[20] J. Ruan, H. Zhang, X. Fang, A. Li, W. C. Campbell, E. Hudson, D. Hayes, H. Haeffner, T. Humble, J. Palsberg *et al.*, Trapsimd: Simd-aware compiler optimization for 2d trapped-ion quantum machines, arXiv:2504.17886.

[21] T. LeBlond, R. S. Bennink, J. G. Lietz, and C. M. Seck, Tiscc: A surface code compiler and resource estimator for trapped-ion processors, in *Proceedings of the SC'23 Workshops of The International Conference on High Performance Computing, Network, Storage, and Analysis* (Association for Computing Machinery, New York, NY, USA2023), pp. 1426–1435.

[22] J. Viszlai, S. F. Lin, S. Dangwal, J. M. Baker, and F. T. Chong, An architecture for improved surface code connectivity in neutral atoms, arXiv:2309.13507.

[23] R. D. Delaney, L. R. Sletten, M. J. Cich, B. Estey, M. I. Fabrikant, D. Hayes, I. M. Hoffman, J. Hostetter, C. Langer, S. A. Moses *et al.*, Scalable multispecies ion transport in a grid-based surface-electrode trap, Phys. Rev. X **14**, 041028 (2024).

[24] C. Ryan-Anderson, J. G. Bohnet, K. Lee, D. Gresh, A. Hankin, J. Gaebler, D. Francois, A. Chernoguzov, D. Lucchetti, N. C. Brown *et al.*, Realization of real-time fault-tolerant quantum error correction, Phys. Rev. X **11**, 041058 (2021).

[25] S. Dasu, B. Criger, C. Foltz, J. A. Gerber, C. N. Gilbreth, K. Gilmore, C. A. Holliman, N. K. Lysne, A. R. Milne, D. Okuno, G. Vittorini, and D. Hayes, Order-of-magnitude extension of

qubit lifetimes with a decoherence-free subspace quantum error correction code, arXiv:2503.22107.

[26] K. Yamamoto, Y. Kikuchi, D. Amaro, B. Criger, S. Dilkes, C. Ryan-Anderson, A. Tranter, J. M. Dreiling, D. Gresh, C. Foltz *et al.*, Quantum error-corrected computation of molecular energies, arXiv:2505.09133.

[27] D. Greenbaum and Z. Dutton, Modeling coherent errors in quantum error correction, Quantum Sci. Technol. **3**, 015007 (2018).

[28] S. J. Beale, J. J. Wallman, M. Gutiérrez, K. R. Brown, and R. Laflamme, Quantum error correction decoheres noise, Phys. Rev. Lett. **121**, 190501 (2018).

[29] J. K. Iverson and J. Preskill, Coherence in logical quantum channels, New J. Phys. **22**, 073066 (2020).

[30] M. E. Beverland, P. Murali, M. Troyer, K. M. Svore, T. Hoefler, V. Kliuchnikov, G. H. Low, M. Soeken, A. Sundaram, and A. Vaschillo, Assessing requirements to scale to practical quantum advantage, arXiv:2211.07629.

[31] T. LeBlond, C. Dean, G. Watkins, and R. Bennink, Realistic cost to execute practical quantum circuits using direct Clifford+T lattice surgery compilation, ACM Trans. Quantum Comput. **5**, 1 (2024).

[32] S. Bravyi, M. Englbrecht, R. König, and N. Peard, Correcting coherent errors with surface codes, npj Quantum Inf. **4**, 55 (2018).

[33] A. S. Darmawan and D. Poulin, Tensor-network simulations of the surface code under realistic noise, Phys. Rev. Lett. **119**, 040502 (2017).

[34] M. Katsuda, K. Mitarai, and K. Fujii, Simulation and performance analysis of quantum error correction with a rotated surface code under a realistic noise model, Phys. Rev. Res. **6**, 013024 (2024).

[35] Á. Márton and J. K. Asbóth, Coherent errors and readout errors in the surface code, Quantum **7**, 1116 (2023).

[36] A. Miller, C. Ostrove, J. Hines, R. Blume-Kohout, K. Young, and T. Proctor, Efficient simulation of Clifford circuits with small Markovian errors, arXiv:2504.15128.

[37] R. S. Bennink, E. M. Ferragut, T. S. Humble, J. A. Laska, J. J. Nutaro, M. G. Pleszkoch, and R. C. Pooser, Unbiased simulation of near-Clifford quantum circuits, Phys. Rev. A **95**, 062337 (2017).

[38] S. Bravyi, D. Browne, P. Calpin, E. Campbell, D. Gosset, and M. Howard, Simulation of quantum circuits by low-rank stabilizer decompositions, Quantum **3**, 181 (2019).

[39] J. M. Pino, J. M. Dreiling, C. Figgatt, J. P. Gaebler, S. A. Moses, M. Allman, C. Baldwin, M. Foss-Feig, D. Hayes, K. Mayer *et al.*, Demonstration of the trapped-ion quantum CCD computer architecture, Nature (London) **592**, 209 (2021).

[40] S. A. Moses, C. H. Baldwin, M. S. Allman, R. Ancona, L. Ascarrunz, C. Barnes, J. Bartolotta, B. Bjork, P. Blanchard, M. Bohn *et al.*, A race-track trapped-ion quantum processor, Phys. Rev. X **13**, 041052 (2023).

[41] M. F. Brandl, A quantum von Neumann architecture for large-scale quantum computing, arXiv:1702.02583.

[42] B. Lekitsch, S. Weidt, A. G. Fowler, K. Mølmer, S. J. Devitt, C. Wunderlich, and W. K. Hensinger, Blueprint for a microwave trapped ion quantum computer, Sci. Adv. **3**, e1601540 (2017).

[43] C. J. Trout, M. Li, M. Gutiérrez, Y. Wu, S.-T. Wang, L. Duan, and K. R. Brown, Simulating the performance of a distance-3 surface code in a linear ion trap, New J. Phys. **20**, 043038 (2018).

[44] Y. Li and S. C. Benjamin, One-dimensional quantum computing with a 'segmented chain' is feasible with today's gate fidelities, npj Quantum Inf. **4**, 25 (2018).

[45] M. Ye and N. Delfosse, Quantum error correction for long chains of trapped ions, arXiv:2503.22071.

[46] D. J. Wineland, C. Monroe, W. M. Itano, D. Leibfried, B. E. King, and D. M. Meekhof, Experimental issues in coherent quantum-state manipulation of trapped atomic ions, J. Res. Natl. Inst. Stand. Technol. **103**, 259 (1998).

[47] D. Kielpinski, C. Monroe, and D. J. Wineland, Architecture for a large-scale ion-trap quantum computer, Nature (London) **417**, 709 (2002).

[48] Our trapped-ion surface code mapping is very similar to the one proposed in Ref. [42], which we were not aware of at the time of writing Ref. [21].

[49] C. Ryan-Anderson (private communication).

[50] C. Baldwin, Quantinuum hardware specifications (2024), https://github.com/CQCL/quantinuum-hardware-specifications.

[51] D. Hayes, D. Stack, B. Bjork, A. Potter, C. Baldwin, and R. Stutz, Eliminating leakage errors in hyperfine qubits, Phys. Rev. Lett. **124**, 170501 (2020).

[52] Besides, SPAM/mid-circuit measurement and reset (MCMR) errors do not seem to substantially contribute to logical error rates in the simulations of Ref. [24].

[53] C. Gidney, Stim: A fast stabilizer circuit simulator, Quantum **5**, 497 (2021).

[54] J. Behrends and B. Béri, The surface code beyond Pauli channels: Logical noise coherence, information-theoretic measures, and errorfield-double phenomenology, arXiv:2412.21055.

[55] H. Qassim, J. J. Wallman, and J. Emerson, Clifford recompilation for faster classical simulation of quantum circuits, Quantum **3**, 170 (2019).

[56] H. Pashayan, O. Reardon-Smith, K. Korzekwa, and S. D. Bartlett, Fast estimation of outcome probabilities for quantum circuits, PRX Quantum **3**, 020361 (2022).

[57] S. Bravyi and D. Gosset, Improved classical simulation of quantum circuits dominated by Clifford gates, Phys. Rev. Lett. **116**, 250501 (2016).

[58] H. Pashayan, J. J. Wallman, and S. D. Bartlett, Estimating outcome probabilities of quantum circuits using quasiprobabilities, Phys. Rev. Lett. **115**, 070501 (2015).

[59] O. Higgott, Pymatching: A Python package for decoding quantum codes with minimum-weight perfect matching, ACM Trans. Quantum Comput. **3**, 1 (2022).

[60] O. Higgott and C. Gidney, Sparse blossom: Correcting a million errors per core second with minimum-weight matching, Quantum **9**, 1600 (2025).

[61] M. Gutiérrez, C. Smith, L. Lulushi, S. Janardan, and K. R. Brown, Errors and pseudothresholds for incoherent and coherent noise, Phys. Rev. A **94**, 042338 (2016).

[62] M. A. Nielsen and I. L. Chuang, *Quantum Computation and Quantum Information* (Cambridge University Press, Cambridge, UK, 2010).

[63] O. Higgott, M. Wilson, J. Hefford, J. Dborin, F. Hanif, S. Burton, and D. E. Browne, Optimal local unitary en-

coding circuits for the surface code, Quantum **5**, 517 (2021).

[64] A. G. Fowler and C. Gidney, Low overhead quantum computation using lattice surgery, arXiv:1808.06709.

[65] D. M. Debroy, L. Egan, C. Noel, A. Risinger, D. Zhu, D. Biswas, M. Cetina, C. Monroe, and K. R. Brown, Optimizing stabilizer parities for improved logical qubit memories, Phys. Rev. Lett. **127**, 240501 (2021).

[66] A. Y. Kitaev, Quantum computations: Algorithms and error correction, Russ. Math. Surv. **52**, 1191 (1997).

[67] Y. R. Sanders, J. J. Wallman, and B. C. Sanders, Bounding quantum gate error rate based on reported average fidelity, New J. Phys. **18**, 012002 (2015).

[68] E. Huang, A. C. Doherty, and S. Flammia, Performance of quantum error correction with coherent errors, Phys. Rev. A **99**, 022313 (2019).

[69] Z. Cheng, E. Huang, V. Khemani, M. J. Gullans, and M. Ippoliti, Emergent unitary designs for encoded qubits from coherent errors and syndrome measurements, PRX Quantum **6**, 030333 (2025).

[70] J. Watrous, Simpler semidefinite programs for completely bounded norms, arXiv:1207.5726.

[71] A. Javadi-Abhari, M. Treinish, K. Krsulich, C. J. Wood, J. Lishman, J. Gacon, S. Martiel, P. D. Nation, L. S. Bishop, A. W. Cross *et al.*, Quantum computing with Qiskit, arXiv:2405.08810.

[72] https://www.energy.gov/downloads/doe-public-access-plan.